\documentclass[11pt,a4paper]{article}

\pdfoutput=1
\usepackage{jheppub}

\usepackage{amssymb}
\usepackage{amsmath}
\usepackage{amsfonts}
\usepackage{graphicx}
\usepackage{color}
\usepackage{xspace}
\usepackage{ulem}
\usepackage{mathtools}
\usepackage{hhline}
\usepackage{float}

\begin{document}

\title{Primordial black hole superradiance and evaporation in the string axiverse}

\author[a]{Marco Calz\`a}\emailAdd{mc@student.uc.pt}
\author[a]{Jo\~ao G.~Rosa}\emailAdd{jgrosa@uc.pt}
\author[a]{Filipe Serrano}\emailAdd{filipeserranophysics@gmail.com}

\affiliation[a]{Univ Coimbra, Faculdade de Ci\^encias e Tecnologia da Universidade de Coimbra and CFisUC, Rua Larga, 3004-516 Coimbra, Portugal.}

\abstract{In the string axiverse scenario, light primordial black holes may spin up due to the Hawking emission of a large number of light (sub-MeV) axions. We show that this may trigger superradiant instabilities associated with a heavier axion during the black holes' evolution, and study the coupled dynamics of superradiance and evaporation. We find, in particular, that the present black hole mass-spin distribution should follow the superradiance threshold condition for black hole masses below the value at which the superradiant cloud forms, for a given heavy axion mass.  Furthermore, we show that the decay of the heavy axions within the superradiant cloud into photon pairs may lead to a distinctive line in the black hole's emission spectrum, superimposed on its electromagnetic Hawking emission.}

\maketitle

\section{Introduction}

There has been a growing interest in the literature on the study of primordial black holes (PBHs), originally predicted by Hawking to form in the early Universe through the direct gravitational collapse of overdense regions \cite{Hawking:1971ei, Carr:1974nx, Carr:1975qj}. These PBHs are natural candidates to account for at least a fraction, and potentially all, the dark matter in the Universe \cite{Clesse:2017bsw, Escriva:2022duf}. Within the standard cosmological paradigm, such PBHs would be born with very little natal spin, given that the ambient radiation pressure would lead to a nearly spherical gravitational collapse. This could explain why the merging BHs recently detected with the LIGO/Virgo/Kagra gravitational wave interferometers seem to be slowly rotating  \cite{Sasaki:2016jop}.

PBHs, particularly light ones with sub-solar masses, also offer new avenues for research in fundamental particle physics, given their ability to produce large numbers of particles, both known and exotic, through Hawking emission \cite{Hawking:1975vcx} and rotational superradiance (see \cite{Brito:2015oca} and references therein). Although in both cases the total number of particles produced over the PBHs lifetime decreases with the PBH mass as $(M/M_P)^2$, where $M_P\simeq 2.176\times10^{-8}$ kg denotes the Planck mass, light PBHs have larger Hawking temperatures $T_H\simeq M_P^2/8\pi M$ and rotate with larger angular velocities $\Omega_H\simeq \tilde{a}/2M(1+\sqrt{1-\tilde{a}^2})$, for a given dimensionless spin parameter $\tilde{a}$. Since Hawking emission and rotational superradiance can only efficiently produce particles of mass $\mu\lesssim T_H$ and $\mu\lesssim\Omega_H$, light PBHs may therefore generate much heavier particles than their stellar or supermassive counterparts. This is appealing since most extensions of the Standard Model predict the existence of exotic particles across many orders of magnitude in mass, well beyond the ``ultra-light regime" accessible with known astrophysical BHs \cite{Arvanitaki:2009fg, Arvanitaki:2010sy, Pani:2012bp, Pani:2012vp, Witek:2012tr, Brito:2013wya, Brito:2014wla, Arvanitaki:2014wva, Arvanitaki:2016qwi, Baryakhtar:2017ngi, Brito:2017wnc, Brito:2017zvb, Cardoso:2018tly, Baumann:2018vus,  Hannuksela:2018izj, Isi:2018pzk, Boskovic:2018lkj, Ikeda:2019fvj, Ghosh:2018gaw, Berti:2019wnn, Baumann:2019eav, Sun:2019mqb, Cannizzaro:2020uap, Brito:2021war, Caputo:2021efm, Cannizzaro:2021zbp, Dias:2023ynv}. 

The timescales involved in Hawking evaporation and superradiance also decrease with the BH mass, so that particle production may efficiently occur at early times or even continuously throughout the cosmic history. PBHs have, in particular, been shown to be relevant for the generation of dark matter \cite{Fujita:2014hha, Allahverdi:2017sks, Lennon:2017tqq, Hooper:2019gtx, Hooper:2020evu, March-Russell:2022zll, Bernal:2022oha}, including both light (but not ultra-light) and heavy axions \cite{Rosa:2017ury, Bernal:2021yyb, Bernal:2021bbv, Calza:2021czr}, and to probe the existence of new particles beyond the energy/luminosity reach of current particle accelerators \cite{Baker:2021btk, Baker:2022rkn}. If they have indeed formed in the early Universe, PBHs may thus provide unique laboratories for fundamental physics.

A particularly interesting setup for PBH particle production is the {\it string axiverse} scenario, which conjectures that realistic string theory compactifications lead to hundreds or even thousands of light axion fields, whose masses are generated only through non-perturbative effects. The number of axion fields is simply dictated by the large number of non-trivial cycles in the six compact extra-dimensions supporting the higher-dimensional Neveu-Schwarz and Ramond-Ramond form-fields, each cycle yielding a pseudo-scalar axion field endowed with a perturbative shift symmetry in the effective 4-dimensional field theory. While axion masses could result from supersymmetry breaking or the mechanism(s) responsible for moduli stabilization, the authors of \cite{Arvanitaki:2009fg} argued that in string compactifications realizing the Peccei-Quinn solution to the strong CP problem, and which therefore include at least one light QCD axion, all axion masses result exclusively from non-perturbative effects. Such scenarios will therefore include a large number of axions spanning a broad range of mass scales.

A population of PBHs born with masses $\sim 10^{12}$ kg will evaporate within the Universe's lifetime. Within the string axiverse scenario, as two of us have shown with March-Russell in \cite{Calza:2021czr}, such PBHs will emit not only photons, electrons and other Standard Model degrees of freedom but also all axions with masses below a few MeV. This changes not only their lifetime (and hence the initial mass of the PBHs that evaporate away before the present day) but also their spin. While the emission of fermions, vector bosons and gravitons inevitably carries away a BH's angular momentum, scalar particles like axions are the only ones that can be emitted in the $l=0$ mode, as originally shown by Taylor, Hiscock and Chambers \cite{Chambers:1997ai, Taylor:1998dk}. Scalar particle emission therefore decreases the BH mass but not its angular momentum, making it spin faster. In \cite{Calza:2021czr} we have simulated the evaporation of light PBHs emitting all known Standard Model particles and an arbitrary number of light axions (mass $\lesssim $ MeV) and, as we review below, we concluded that slowly-rotating PBHs may develop spin parameters $\tilde{a}\gtrsim 0.1$ before evaporating away in the presence of a few hundred light axions. The present mass-spin distribution of light PBHs (at different stages of the evaporation process) depends on the total number of light axions, thus providing a unique probe of the string axiverse. This has the appeal of being a purely gravitational probe of this scenario, independent of how the individual axions interact with known particles, as well as of the details of the axion mass spectrum.

The fact that PBHs naturally develop non-negligible spin parameters through Hawking emission in this scenario motivates exploring whether this may trigger superradiant instabilities. In particular, in the string axiverse spectrum there may also exist a number of {\it heavy axions} ($\mu\gtrsim$ MeV), since the non-perturbative nature of the axion mass generation mechanism only implies that their masses are exponentially suppressed compared to a high mass scale such as the supersymmetry breaking scale. Hence, if the PBHs are born with low spin, the condition for superradiant particle production $\mu<\Omega_H$ will only be satisfied once the PBHs evaporate sufficiently and spin up due to the emission of {\it light axions}.

In this work, we thus study the dynamical generation of superradiant heavy axion clouds around PBHs born with mass $\sim 10^{12}$ kg throughout the cosmic history, including both superradiance and Hawking emission. We will show that indeed such clouds may form with two important observational consequences. 

First, the formation of superradiant clouds spins down the PBHs faster than evaporation can spin them up. This modifies the present PBH mass-spin distribution such that the lightest PBHs (which have evaporated sufficiently for superradiant clouds to form) saturate the superradiance condition, $\Omega_H\simeq \mu$, while the spins of the heavier PBHs are determined solely by their evaporation stage and, hence, by the number of light axion species.

Second, the decay of the heavy axions into photon pairs leads to a characteristic gamma-ray line in the PBH-axion cloud photon emission spectrum. This is a very unique signature since the Hawking emission spectrum (including both primary and secondary photons) evolves as the PBH evaporates, while the line has a fixed energy corresponding to approximately half of the heavy axion's mass.

We begin our discussion by reviewing PBH evaporation through Hawking emission in the next section, discussing in particular the string axiverse case. In section 3 we start by reviewing the dynamics of superradiant instabilities for massive scalar fields and then bring together these two particle production mechanisms by fist considering a toy model where a BH evaporates by emitting a single light axion and a superradiant instability is induced by another heavy axion. Although unrealistic, this toy model allows one to understand the basic dynamics of the problem towards exploring a more realistic setup where all Standard Model particles are included in the PBH evaporation process alongside an arbitrary number of light axions. In section 4, we compute the PBH photon emission spectrum including primary and secondary Hawking emission as well as heavy axion decay within the superradiant clouds. We summarize our main results and conclusions in the final section.

We note in advance that axion self-interactions are assumed to play a negligible role in the dynamical evolution. This is a good approximation for the large axion decay constants typically predicted in string constructions and that we also take into account when discussing observational prospects in section 4.

\section{Hawking emission and black hole evaporation}

In curved space-time, different observers do not necessarily agree in their definition of what is the quantum vacuum state, i.e.~the state with the lowest possible energy. This is due to the use of different time coordinates to perform the separation between the positive and negative frequency modes that underlies the field quantization procedure. While in flat Minkowski space all inertial observers perform this mode separation in an equivalent way due to Lorentz invariance, in space-time manifolds that include regions with non-negligible curvature, particularly event horizons, this is typically not the case.

In particular, Hawking showed in 1974 \cite{Hawking:1974rv,Hawking:1974sw} that a stationary (and therefore non-inertial) observer standing far away from a BH horizon will measure an outgoing flux of particles with a nearly thermal spectrum, if the associated quantum field is in the vacuum state as defined by an observer freely falling into the BH, or equivalently the vacuum state as defined in the asymptotic past well before the collapsing matter formed the BH. This Hawking radiation is therefore a purely gravitational effect, such that a BH essentially emits all particle species with masses below its Hawking temperature, $T_H\simeq M_P^2/8\pi M$ for a slowly rotating BH. These remove the mass and angular momentum of the BH (as measured by an asymptotic observer), thus leading to its evaporation. 

In this section we will compute in detail how light PBHs evaporate by emitting not only the known Standard Model particles but also a large number of scalar axions, as mentioned above, and we are particularly interested in describing the evolution of its mass and spin. This necessarily involves a precise numerical calculation of the spectrum of emitted particles with different spin, namely the associated ``gray-body'' factors that quantify the deviations from a purely Bose-Einstein or Fermi-Dirac distribution. These are a consequence of the effective potential probed by the field modes propagating in the BH space-time, which we will assume to be described by the Kerr solution (since any primordial electric charge is radiated away well before it loses a significant amount of mass or angular momentum \cite{Gibbons:1975kk}). These gray-body factors are, in fact, related to the transmission coefficients $\Gamma^s_{l,m}$ for the associated wave scattering problem in the same BH effective potential, and which can be obtained by solving the radial Teukolsky equation \cite{Teukolsky:1972my, Teukolsky:1973ha, Press:1973zz, Teukolsky:1974yv} describing massless waves of arbitrary spin $s$. We will use a shooting method to numerically solve this equation (see e.g.~\cite{Rosa:2016bli}). We will then use these results to study the dynamical evolution of a rotating black-hole using the formalism described in \cite{Page:1976df,Page:1976ki,Page:1977um,Chambers:1997ai,Chambers:1997ax,Taylor:1998dk}.

For simplicity, in this section we consider geometrized units such that $\hbar=c=G=1$ ($M_P=1$).


\subsection{Quantum fields in the Kerr space-time}

We consider a rotating black hole described by the Kerr solution, which in Boyer-Lindquist coordinates $(t,r,\theta,\varphi)$ reads
\begin{align} \label{Kerr_metric}
    &ds^2 = \left( 1- \frac{2 M r}{\Sigma}\right)dt^2 + \frac{4 M a r \sin^2 \theta }{\Sigma}dt d\varphi + \\ \nonumber
    &-\frac{\Sigma}{\Delta}dr^2 - \Sigma d\theta^2 - \left(r^2 + a^2 + \frac{2 M a^2 r \sin^2 \theta}{\Sigma}\right) \sin^2 \theta d\varphi^2,
\end{align}
where $M$ is the hole mass, $a=J/M$ is the BH angular momentum, $\Delta=r^2 + a^2 - 2 M r$, and $\Sigma=r^2 a^2 \cos^2 \theta$.
This solution has an inner Cauchy horizon and an outer event horizon at $r=r_\pm$ ($\Delta(r_\pm)=0$).

Despite the different equations governing the dynamics of massless (test) fields in curved space-time, in the case of the Kerr metric a unified description can be obtained with the aid of the Newman-Penrose (NP) formalism \cite{Pirani1964,Kinnersley:1969zza,Newman:1961qr}, and one can condense all these equations into the so-called Teukolsky master equation, which in Boyer-Lindquist coordinates is given by:
\begin{align}
    &\frac{1}{\Delta^s} \partial_r (\Delta^{s+1} \partial_r \Upsilon_s) + \frac{1}{\sin \theta } \partial_\theta (\sin \theta \partial_\theta \Upsilon_s) - \left(\frac{(r^2+a^2)^2}{\Delta} - a^2 \sin^2 \theta \right)\partial^2_t + \nonumber \\
    &-\frac{4 M a r }{\Delta}\partial_t \partial_\varphi \Upsilon_s - \left(\frac{a^2}{\Delta} - \frac{1}{\sin^2 \theta }\right) \partial^2_\varphi \Upsilon_s+ 2 s \left(\frac{M(r^2 - a^2)}{\Delta} -r -i a \cos \theta\right) \partial_t \Upsilon_s+ \nonumber \\
    &+ 2 s \left( \frac{a(r-M)}{\Delta} + \frac{i \cos \theta }{\sin^2 \theta }\right) \partial_\varphi \Upsilon_s - (s^2 \cot^2 \theta -s) \Upsilon_s=0~, \label{Teq}
\end{align}
where the functions $\Upsilon_s$ encode the NP scalars, obtained by contraction of the original tensor fields with the Kinnersley tetrad null vectors \cite{Kinnersley:1969zza}. This equation thus encodes the dynamics of Klein-Gordon scalar fields ($s=0$), Weyl fermions ($s=\pm 1/2$), vector fields ($s=\pm1$), Rarita-Schwinger fields ($s=3/2$) and gravitational perturbations ($s=\pm 2$) in the massless limit. Moreover, the Teukolsky equation can be solved by separation of variables, with the NP scalars admitting a mode decomposition of the form:
\begin{equation}
    \Upsilon_s= \sum e^{-i \omega t } e^{i m \varphi} S_s(\theta) R_s(r)~,
\end{equation}
where $\omega$ is the perturbation frequency, $m$ is the azimuthal angular momentum quantum number. The angular functions $S_s(\theta)$ are the so-called spin-weighted spheroidal harmonics, which satisfy the equation \cite{Fackerell&Crossman1977,Suffern1983,Seidel:1988ue,Berti:2005gp} 
\begin{align}\label{SWSH}
    &\frac{1}{\sin \theta} \partial_\theta \left( \sin \theta \textcolor{white}{a}  \partial_\theta S_s\right) + \left( a^2 \omega^2 \cos^2 \theta - \frac{m^2}{\sin^2 \theta } -2 a \omega s \cos \theta \right. + \\ \nonumber
    &\left. -\frac{2 m s \cos \theta }{\sin^2 \theta} -s^2 \cot^2 \theta +s+  {_s A^m_l }\right) S_s=0~.
\end{align}
These functions reduce to scalar spherical harmonics for $s=0$ and $a=0$, and generalize conventional spin-weighted spherical harmonics to the more general case of axial-symmetry.
${_s A^m_l } = {_s A^m_l }(a\omega)$ are the eigenvalues of (\ref{SWSH}) and cannot be expressed analytically in terms of the spherical angular momentum quantum numbers $l, m$. Nevertheless, for $a \omega \ll 1$ they can be computed using a perturbative expansion, yielding:
\begin{eqnarray}
{}_s A^m_l (a\omega) &=&l(l+1) - s(s+1) -a\omega \frac{2ms^2}{l(l+1)}\nonumber\\
&+& (a\omega)^2 \left\{\frac{2}{3} \left[ 1+ \frac{3m^2-l(l+1)}{(2l-1)(2l+3)}\right] - \frac{2s^2}{l(l+1)} \frac{3m^2-l(l+1)}{(2l-1)(2l+3)} \right.\nonumber\\
&+& \left.2s^2 \left[ \frac{(l^2-s^2)(l^2-m^2)}{l^3(2l-1)(2l+1)} - \frac{[(l+1)^2-m^2][(l+1)^2 - s^2]}{(l+1)^3(2l+1)(2l+3)}\right]\right\} +\mathcal{O}\left[(a \omega)^3\right]~.
\end{eqnarray}
The functions $R_s$ are the radial part of the NP scalars and satisfy the radial equation
\begin{equation}\label{radTeq}
\Delta^{-s} \partial_r( \Delta^{s+1}\partial_rR_s) + (K^2 -2is(r-M)K)\Delta^{-1}+4 i s \omega r - _s Q^m_l)R_s=0~,
\end{equation}
where $_s Q^m_l={_s A^m_l }+ a^2 \omega^2 - 2 a \omega m$ and $K=(r^2 + a^2) \omega - m a  $. These functions take the following form far away and near the BH horizon:
    \begin{align}
   & R_s \sim R^{in}_s \frac{e^{-i\omega r_*}}{r}+ R^{out}_s \frac{e^{i\omega r_*}}{r^{2s+1}}~, \quad r\gg r_+ \\ 
    & R_s \sim R^{hole}_s \Delta^{-s} e^{-i k r_*}~, \quad r-r_+\ll r_+
    \end{align}
where $r_*$ is the tortoise coordinate, defined via $dr_*/dr=(r^2+a^2)/\Delta$, and we have imposed ingoing boundary conditions at the horizon.
We note that the solutions $R_s$ and $R_{-s}$ for the same spin are, in general, distinct, but are nevertheless related through the Teukolsky-Starobinsky identities \cite{Starobinskii&Churilov1973,Mano:1996vt,Mano:1996mf,Mano:1996gn,Fiziev:2009ud} that can be derived from the original field equations:
\begin{align}
    &\Delta^s(\mathcal{D^\dagger})^{2s}  \Delta^s R_s = C^*_s R_{-s}~, \\
    &(\mathcal{D})^{2s} R_{-s}= C_s R_{s}~,
\end{align}
where $\mathcal{D}=\partial_r - i \frac{K}{\Delta}$. The Starobinsky constants $C_s$ (omitting the $l,m$ quantum numbers for simplicity) for the fields of interest are given by: 
\begin{align}
    &C_0^2=1 \\
    &C_{1/2}^2=Q_{1/2}+ \frac{1}{4} \\
    &C_{1}^2=Q_{1}^2+ -4 a^2 \omega^2 -2 a \omega m \\
    &C_{3/2}^2=\left(Q_{3/2}+ \frac{3}{4}\right) \left(Q_{3/2} + \frac{1}{4}\right)- 16 a^2 \omega^2 \left(Q_{3/2} - \frac{7}{4}\right) + 16 a m \omega \left(Q_{3/2} - \frac{3}{4}\right) \\
    &|C_2|^2=(Q_2^2 + 4 a \omega m - 4 a^2 \omega^2 )((Q_2-2)^2 + 36 a \omega m - 36 a^2 \omega^2 ) + \\ \nonumber
    &+ (2 Q_2-1)(96a^2 \omega^2 - 48 a \omega m ) + 144 \omega^2 (M^2 - a^2)~.
\end{align}
Note that $C_0, C_{1/2}, C_{1},$ and $C_{3/2}$ are real, while $C_2$ is imaginary.\\

Through the redefinition $Y_s= \Delta^{s/2}(r^2+ a^a)^{1/2}R_s$ we may also write (\ref{radTeq}) in a Schr\"odinger-like form:
\begin{equation}\label{S-leq}
    (\partial^2_{r_*}-V)Y_s=0~,
\end{equation}
where the effective potential $V=$$_s V^m_l$ vanishes both at the horizon, where $r_* \rightarrow - \infty$, and at infinity, for $r_*
\rightarrow + \infty$.
The form of Eq.~(\ref{S-leq}) guarantees that the Wronskian does not change if calculated at different radial positions. In particular, there is a  conserved current:
\begin{equation}\label{wrons}
    \left[Y^*_{-s} \partial_{\tilde{r}}Y_s - Y_{s} \partial_{\tilde{r}}Y^*_{-s}\right ]_{r=r_+}=\left[Y^*_{-s} \partial_{\tilde{r}}Y_s - Y_{s} \partial_{\tilde{r}}Y^*_{-s} \right]_{r=\infty}.
\end{equation}
Substituting the asymptotic and near-horizon solutions in the Teukolsky-Starobinsky identities yields the relations between $R_{+|s|}$ and $R_{-|s|}$, while (\ref{wrons}) yields an energy conservation law, with equal energy flux at the horizon and at infinity.

In the wave scattering problem, the transmission coefficient is then given by the ratio between the energy flux into the BH horizon and the incoming energy flux at infinity:
\begin{equation}
    \Gamma= \frac{dE_{hole}/dt}{dE_{in}/dt}~.
\end{equation}
Note that this coefficient depends on the frequency, spin and angular momentum quantum numbers of each field mode, as well as on the BH spin parameter, $\Gamma= {_s\Gamma ^l_m }(a , \omega)$. One can show that the same coefficient yields the corresponding gray-body factor for Hawking emission, since in the latter case it quantities the filtering of field modes by the BH effective potential as they propagate away from the event horizon.

\subsection{Numerical computation of gray-body factors}

An analytical computation of the transmission coefficients is only possible under very stringent approximations \cite{Starobinsky:1973aij}, so numerical methods are in general required to compute them for different wave modes. Here, we will use a shooting method similar to the one employed in e.g.~\cite{Rosa:2016bli} and first proposed by Starobinsky \cite{Starobinsky:1973aij}.
The first step is to write Eq.~(\ref{radTeq}) in terms of the re-scaled radial coordinate $x=(r-r_+)/r_+$:
\begin{equation}\label{Teuk2}
    x^2(x+\tau)^2 \partial_x^2 R(x) +(s+1)(2 x+\tau)x(x+\tau) \partial_x R(x) + V(x)R(x)=0~,
\end{equation}
where the effective radial potential can be written as:
\begin{equation}
    V(x)=k^2- i s (2 x + \tau) k + (4 i s \omega (x+1)-Q^{l,m}_s )x (x+\tau)~,
\end{equation}
with $k=(2-\tau)(\omega-m \Omega_H) r_+ + x(x+2)\omega r_+$, and $\tau=(r_+ - r_-)/r_+$. Imposing ingoing boundary conditions at the horizon, the near-horizon solutions of Eq.~(\ref{Teuk2}) can then be expressed in a Taylor expansion \cite{Rosa:2016bli, Rosa:2012uz} of the form
\begin{equation}\label{near}
    R(x)= x^{-s- i \varpi/\tau} \sum_{n=0}^\infty a_n x^n,
\end{equation}
where $\varpi=(2-\tau)(\omega-m \Omega_H) r_+ $ and the coefficients $a_n$ can be determined by substituting the power series (\ref{near}) in (\ref{Teuk2}) and solving iteratively the resulting algebraic equations. The near-horizon solution is then used as a boundary condition for numerically integrating the radial Teukolsky equation up to large distances, where the general form of the solution is known and reads:
\begin{equation}
    R(x) \rightarrow \frac{_s R^{l m }_{in}}{r_+} \frac{e^{-i \bar \omega x}}{x} + \frac{_s R^{l m }_{out}}{r_+^{2s+1}} \frac{e^{i \bar \omega x}}{x^{2s+1}}~,
\end{equation}
where $\bar \omega = \omega r_+ $.
It is then possible to extract the coefficient $_s R^{l m}_{in}(\omega)$ in order to evaluate the transmission coefficient.
The normalization of the scattering problem is set by setting e.g.~$a_0=1$ which is equivalent to 
\begin{equation}  
    | _s R^{l m }_{hole}|^2=(2 r_+ )^{2s}(a^2 -M^2)^2~.
\end{equation}
This then yields the transmission coefficients for the different spin fields:
\begin{align}
     &\Gamma^{l m}_0= | _0 R^{l m}_{in}|^{-2} \\
     &\Gamma^{l m}_{1/2}=  \tau| _{1/2} R^{l m}_{in}|^{-2}\\
     &\Gamma^{l m}_1(\omega)= \frac{\tau^2  \bar\omega }{\varpi}| _1 R^{l m}_{in}|^{-2}\\
     &\Gamma^{l m}_{3/2}=  \frac{\tau^3 {\bar\omega}^2}{(\varpi^2 + \tau^2/16)} | _{3/2} R^{l m}_{in}|^{-2}\\
     &\Gamma^{l m}_2=  \frac{\tau^4 {\bar\omega}^3}{\varpi(\varpi^2 + \tau^2/4)} | _2 R^{l m}_{in}|^-{2} ~,
\end{align}
or, in a more compact way,
\begin{equation}
   \Gamma^{l m}_s=\delta_s | _s R^{l m}_{in}|^{-2}
\end{equation}
with
\begin{equation}
\delta_s=- i e^{i \pi s} \omega^{4 s - 2} \left(\frac{\tau}{4}\right)^{1-2s} \frac{\Gamma(1-s+i4\frac{\eta}{\tau})}{\Gamma(s+i4\frac{\eta}{\tau})} ~.
\end{equation}

\subsection{Hawking evaporation in the string axiverse}

We determine the evolution of PBHs following the formalism described in \cite{Page:1976df,Page:1976ki,Page:1977um} and later in \cite{Chambers:1997ai,Chambers:1997ax,Taylor:1998dk}. The PBH mass and spin evolution is determined by the functions $\mathcal{F}\equiv -M^2 dM/dt$ and $\mathcal{G}\equiv-(M/\tilde a) dJ/dt$, which remove the dependence on the BH mass. Here $\tilde a = a/M$ is the BH dimensionless spin parameter. These are given by:
\begin{equation}\label{f_g}
\begin{pmatrix}
\mathcal{F}\\
\mathcal{G}
\end{pmatrix}=\sum_{i,l,m}\frac{1}{2\pi} \int_0^{\infty}dx \frac{^s\Gamma_{i,l,m}}{e^{2\pi k/\kappa}\pm 1}
\begin{pmatrix}
x\\
m \tilde{a}^{-1}
\end{pmatrix}~,
\end{equation}
where the sum is taken over all particle species $i$ and angular momentum quantum numbers $(l,m)$, $x=\omega M$, $k=\omega-m\Omega_H$ and $\kappa=\sqrt{1-\tilde{a}^2}/2r_+$ is the surface gravity of the Kerr BH, with $\Omega_H$ denoting the angular velocity at the event horizon, located at $r_+$. The upper/lower sign corresponds to fermion/boson fields. The function
\begin{equation}
  \mathcal{H}=\frac{\mathcal{G}}{\mathcal{F}}-2
\end{equation}
determines whether a black hole spins up or down during its evolution, taking into account the relative magnitude of mass and angular momentum loss rates. If there is a value $\tilde{a}_*$ for which $\mathcal{H}(\tilde{a}_*)=0$, the PBH spin parameter will tend to this stable value provided that $\partial_{\tilde{a}} \mathcal{H}|_{\tilde{a}=\tilde{a}_*}>0$. We note that for $\partial_{\tilde{a}} \mathcal{H}|_{\tilde{a}=\tilde{a}_*}\leq0$ the equilibrium point is unstable but that we will not find such cases in our analysis. 
The differential equations governing the PBH spin and mass evolution can be written in terms of dimensionless variables useful for numerical integration:
\begin{equation}
    y=-\ln{\tilde a},\qquad z=-\ln{M/M_i},
   \qquad \tau=-M^{-3}_i t~,
\end{equation}
such that
\begin{equation}
    z'(y)=\frac{1}{\mathcal{H}}~,
\qquad
    \tau'(y)=\frac{e^{-3z(y)}}{\mathcal{H} \mathcal{F}}~,
\end{equation}
where $M_i$ is the initial BH mass, with initial conditions $z(t=0)=0$ and $\tau(t=0)=0$.

The numerical method described in section 2.2 allow us to compute the gray-body factors for massless fields. In principle one may compute these for massive fields, but given that the emission of particles with masses, $\mu$, above the Hawking temperature $T_H =\kappa/2\pi \simeq 1\, {\rm GeV} (10^{10}{\rm kg}/M)$ is exponentially suppressed we work in the approximation where particles are considered massless for $T_H>\mu$ and are otherwise absent from the emission spectrum.
Massless particles as photons, gravitons are emitted since the BH forms alongside all particles with mass $\mu<T_H \sim $ few MeV like neutrinos and electrons/positrons, given that this is the natal temperature of PBHs with a lifetime comparable to the age of the Universe, for which $M_i\sim 10^{12}$ kg.

As a PBH evaporates its Hawking temperature increases, allowing for the emission of more and more massive degrees of freedom, like muons, tau particles, etc, above the corresponding mass thresholds. The only hadrons with mass below the QCD scale are pions ($\pi^0$ and $\pi^{\pm}$), these being the only hadronic states included directly in the BH emission spectrum. Temperatures above the QCD scale allow for the direct emission of elementary quarks and gluons that subsequently hadronize. Following \cite{Halzen:1990ip,MacGibbon:1990zk,Halzen:1991uw,MacGibbon:1991tj,MacGibbon:1991vc,MacGibbon:2007yq,Ukwatta:2009xk,MacGibbon:2010nt,MacGibbon:2015mya,Ukwatta:2015iba}, we have considered the effective quark and gluon QCD masses given in \cite{ParticleDataGroup:2006fqo}, taking these as threshold values above which each particle is included in the PBH emission spectrum. We note that our results do not change significantly if we consider other values for the effective quark and gluon masses given in the literature, such as in \cite{Iritani:2009mp}.

\begin{figure}[htbp]
\centering\includegraphics[width=0.65\columnwidth]{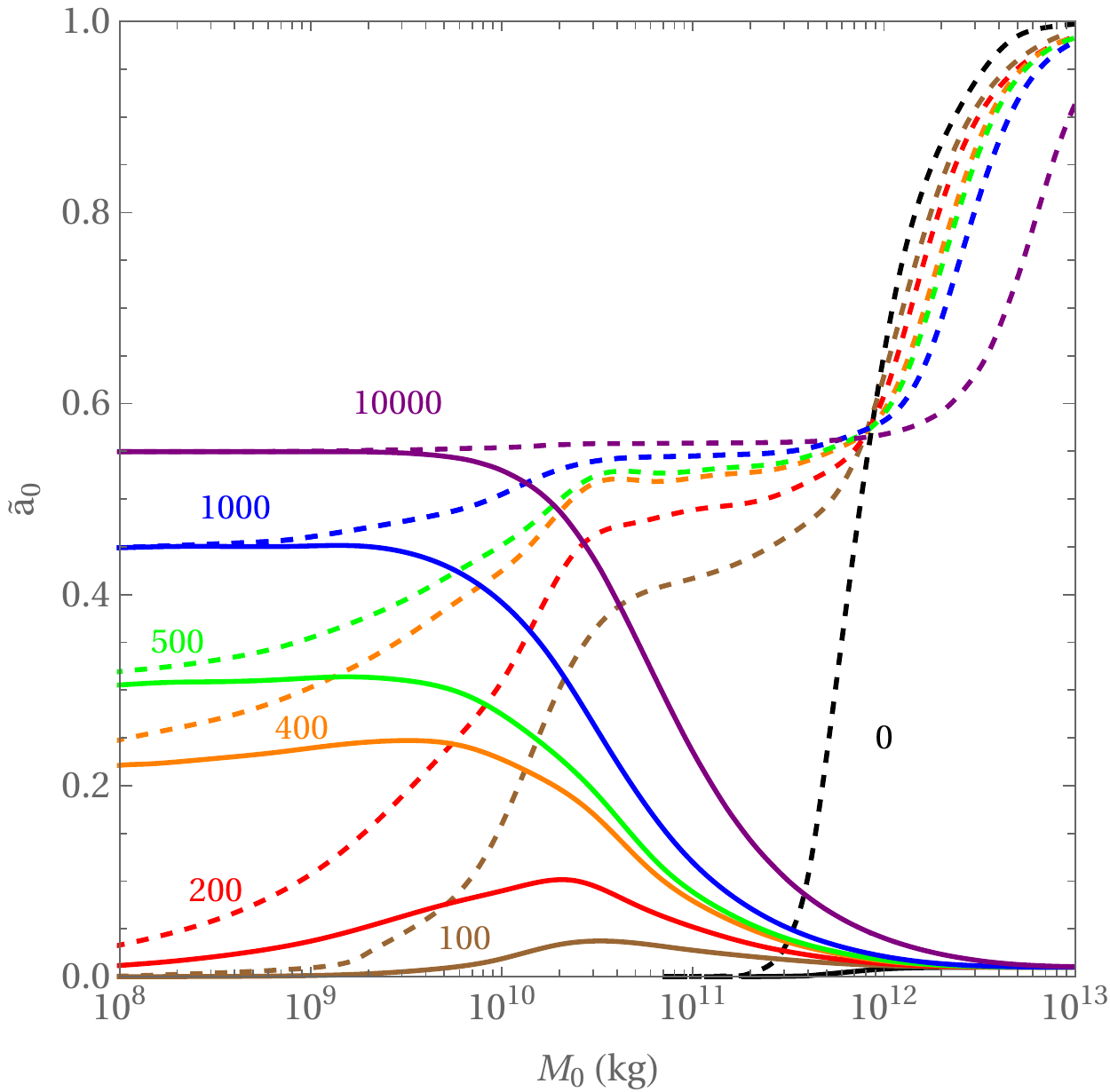}
\caption{Present PBH spin, $\tilde{a}_0$, as a function of the corresponding present mass, $M_0$, for an initial population with spin $\tilde{a}_i=0.01$ (solid curves) or $\tilde{a}_i=0.99$ (dashed curves). Curves are labelled (and coloured) according to the number of light axions ($\lesssim$ few MeV) included in the PBH emission spectrum.}\label{fig1}
\end{figure}

In order to reproduce the results first obtained in \cite{Calza:2021czr}, in addition to the Standard Model particles we have considered an arbitrary number of light axions, $N_a$, corresponding to the fraction of the string axiverse with mass below a few MeV. We show, in Fig.~\ref{fig1}, our results for the present spin of PBHs, $\tilde{a}_{0}$, as a function of their present mass, $M_0$, for different numbers of axions. We consider two limiting cases for the natal PBH spin: $\tilde{a}_{i}=0.01$ (solid curves) and $\tilde{a}_{i}=0.99$ (dashed curves), corresponding to PBH formation in the radiation-dominated era \cite{Chiba:2017rvs, Mirbabayi:2019uph, DeLuca:2019buf, Harada:2020pzb} or in an early matter-dominated era \cite{Harada:2017fjm}, respectively. The reader should note that the PBHs of Fig.~\ref{fig1} correspond to the remnants of an initial population of PBHs with nearly the same mass and which are presently at different stages of their evolution, justifying the assumption of a common initial spin. This means that Fig.~\ref{fig1} also depicts the time evolution of the PBH spin, with time flowing from right to left. Note also that the initial mass of PBHs with a lifetime matching the age of the Universe of 13.8 Gyrs depends on the number of emitted species, in particular the number of light axions. In particular, this critical initial mass ranges from $5\times10^{11}\,$kg in the absence of axions to $\sim 2.7\times10^{12}\,$kg for $N_a=1000$, scaling as $N_a^{1/3}$ for $N_a\gtrsim 10$.

As one can see in this figure, in the absence of axions ($N_a=0$, black curves) PBHs lose their spin quite quickly, such that any PBHs with present mass $\lesssim10^{11}$ kg should have negligible spin, i.e.~spin parameters well below the percent level. A drastic change in this picture occurs in the string axiverse for $N_a\gtrsim 100$, with PBHs initially spinning up due to the emission of a large number of light scalars (or spin loss being initially halted for initially near-extremal PBHs). When the PBH mass approaches $\sim 10^{10}$ kg and the corresponding Hawking temperature exceeds the QCD scale, the large number of spin-1/2 and spin-1  degrees of freedom emitted starts counteracting the light scalar emission, effectively spinning down the PBH as it evaporates for $N_a\lesssim 400$. Above this number of light axions, the PBH spin asymptotes to a non-vanishing value, which tends to the critical value $\tilde {a}\simeq 0.555$ originally found in \cite{Chambers:1997ai} for pure scalar emission as $N_a \rightarrow \infty$.

Fig.~\ref{fig1} shows that, independently of their natal spin, PBHs with present mass $M_0 \lesssim 10^{11}$ kg should have a non-negligible spin in the string axiverse scenario (for $N_a\gtrsim 100$). For the case of initially slowly spinning PBHs, this is particularly relevant, since this spin up due to Hawking emission may render them unstable with respect to superradiant particle creation, namely if the string axiverse includes (as one may expect) heavier axions. As we will analyze in the next sections, this may have a dramatic effect on the PBH spin evolution, making the present PBH mass-spin distribution an even powerful probe of the string axiverse spectrum, with potential directly observable signatures.

\section{Superradiant instabilities for evaporating PBHs}

\subsection{Basics of black hole superradiance}

Before analyzing how superradiant instabilities may be triggered by PBH evaporation, as suggested by the analysis of \cite{Calza:2021czr} reviewed in the previous section, we begin by discussing the basic dynamical features of black hole superradiance neglecting the effects of Hawking emission. Consider then a massive scalar field minimally coupled to gravity, of mass $\mu$, described the action:
\begin{equation}
    \centering
    S = \int d^4x \sqrt{-g} \ \left(\frac{1}{2}g^{\mu\nu}\partial_\mu\Phi\partial_\nu\Phi - \frac{1}{2} \mu^2 \Phi^2\right)~,
    \label{actionphi}
\end{equation}
from which we may derive the corresponding equation of motion in the Kerr metric Eq.~(\ref{Kerr_metric}):
\begin{eqnarray}  \label{phii}
&&\frac{\left( r^2+a^2 \right)^2-a^2\sin^2\theta}{\Delta\Sigma}\partial^2_t\Phi-\frac{1}{\Sigma}\partial_r\left( \Delta\partial_r\Phi \right)-\frac{1}{\Sigma\sin\theta}\partial_\theta\left( \sin\theta\partial_\theta\Phi \right)\nonumber\\
&-&\frac{\Delta-a^2\sin^2\theta}{\Sigma\Delta\sin^2\theta}\partial^2_\varphi\Phi  
    +\frac{2a}{\Delta\Sigma}2Mr\partial_t\partial_\varphi\Phi +\mu^2\Phi=0~,
\end{eqnarray}
which reduces to the Teukolsky equation (\ref{Teq}) for $s=0$ in the massless limit. Similarly to the latter, the massive Klein-Gordon equation admits a mode decomposition of the form:
\begin{equation}
    \Phi(t,r,\theta,\varphi) = R_{n,l}(r)S_{l,m}(\theta)e^{-i\omega t}e^{im\varphi}~,
    \label{ansatz}
\end{equation}
where $S_{l,m}(\theta)e^{im\varphi}$ denote scalar spheroidal harmonic functions and  now the radial function, $R_{n,l}(r)$ obeys a ``massive'' Teukolsky equation
\begin{equation}
    \centering
    \Delta\partial_r(\partial_r R_{n,l}) - \Delta\big[\mu^2r^2 + a^2\omega^2 - 2\omega m a r + \big(\omega(r^2+a^2)-ma\big) + \lambda \big] R_{n,l} = 0~.
\end{equation}
It is well known that this equation admits quasi-bound state solutions with complex frequencies $\omega =\omega_R + i\omega_I$, where $\omega_R<\mu$ such that the field is trapped in the gravitational potential well created by the BH \cite{Damour:1976kh, Zouros:1979iw, Detweiler:1980uk, Furuhashi:2004jk, Cardoso:2005vk, Dolan:2007mj, Rosa:2009ei, Rosa:2012uz, Dolan:2012yt, Brito:2015oca, East:2017ovw, East:2017mrj, Dolan:2018dqv}. In the non-relativistic limit, where the dimensionless mass coupling
\begin{equation}
  \alpha={\mu M\over M_P^2}  
\end{equation}
is small, the real part of the quasi-bound state spectrum approaches a Hydrogen-like form
\begin{equation}
\centering
\omega_R = \mu \left(1-\frac{\alpha^2}{2n^2}\right)
\end{equation}
where $\alpha$ plays the role of the fine-structure constant, a simple consequence of the fact that, when written in the Schr\"odinger-like form Eq.~(\ref{S-leq}) the potential is essentially Coulomb-like at large distances from the event horizon (where the scalar field finds support in the $\alpha\lesssim 1$ regime), $V(r)\simeq -\alpha/r$. As for an electron in a Hydrogen atom, the typical velocity is $\sim \alpha$, which justifies denoting $\alpha\ll 1$ as the non-relativistic regime.

The imaginary part, $\omega_I$, reflects the instability of the bound-states (hence the use of the prefix ``quasi-''), with $\omega_I<0$ corresponding to a decay or absorption of the scalar field by the BH, and  $\omega_I>0$ to an exponential amplification of the field and of the associated particle number. For $\alpha\ll 1$, one finds an approximate analytical expression for the imaginary part of the frequency given by:
\begin{align}
    \omega_I&=-\frac{1}{2}\Big(\frac{l!}{(2l+1)!(2l)!}\Big)^2 \frac{(l+n)!}{(n-l-1)!}\frac{4^{2l+2}}{n^{2l+4}}\times\nonumber\\
    &\times\prod_{k=1}^l\Big(k^2+16\Big(\frac{M(\omega_R - m\Omega_H)}{\tau}\Big)^2\Big)\Big(\frac{\omega_R - m\Omega_H}{\tau}\Big)\alpha^{4l+5}\Big(\frac{r_+-r_-}{r_++r_-}\Big)^{2l+1}  \label{eq: 34}
\end{align}
The {\it superradiant instability} thus occurs whenever $\omega_R<m \Omega_H$ ($\alpha<\tilde{a}/4$ for slowly spinning BHs), leading to an extremely efficient production of particles forming a bound {\it superradiant cloud} around the spinning BH. We may regard this as a kind of stimulated emission (even though the process is classical) since all produced particles have the same quantum numbers. In particular, for the fastest growing ``2p-state'' ($n=2$, $l=m=1$):
\begin{equation}
    \centering
    \omega_I = -\frac{1}{12}\Big(1+16\Big(M\frac{\omega_R-\Omega_H}{\tau}\Big)^2\Big)\Big(\frac{\omega_R - \Omega_H}{\tau}\Big)\alpha^{9}\Big(\frac{r_+-r_-}{r_++r_-}\Big)^{3} 
    \label{omega_I}
\end{equation}
Note that the number of particles grows twice as fast, since $N\propto \Phi^2$. Also, the ``2p-state'' grows exponentially faster than all others, so we may neglect any other modes in the dynamics of superradiance (neglecting self-interactions, as we discuss below). Thus, each particle produced by superradiance carries one unit of spin from the BH, along side a mass $\mu$, so energy and angular momentum conservation yield:
\begin{equation}
    \centering 
    \frac{dM}{dt} =-\mu \frac{dN}{dt}~,  \qquad \frac{dJ}{dt} =-\frac{dN}{dt}~, 
    \label{EOMm}
\end{equation}
such that the dimensionless spin parameter evolves according to:
\begin{equation}
 \frac{d\tilde{a}}{dt} = -\frac{M_P^2}{M^2}(1-2\tilde{a}\alpha)\frac{dN}{dt}\simeq -\frac{M_P^2}{M^2}\frac{dN}{dt}~,
    \label{EOMj}
\end{equation}
where in the last step we considered the limit $\alpha\ll 1$. The number of particles within the superradiant cloud then follows:
\begin{equation} \label{dNdt}
    \centering 
\frac{dN}{dt} = \Gamma_s(M,\tilde{a},\mu) N
\end{equation}
where $\Gamma_s = 2\omega_I$. It will be useful to note that, for slowly rotating BHs: 
\begin{equation} \label{Gamma_s}
    \centering
    \Gamma_s \simeq \frac{1}{24}(\tilde{a} - 4\alpha)\alpha^8\mu 
\end{equation}
Note that, strictly speaking, the instability growth rates are computed assuming a fixed BH mass and spin parameter, but since $\mu\ll M$ and $M\gg M_P$ in the regime of interest to our discussion, we may take this a good approximation. The same is true for the semi-classical calculation of the Hawking emission rate, and we may for similar reasons take the two particle production processes as independent, specially since they typically involve different particle species as we discuss below.

\subsection{A toy model}

Given the discussion in the previous subsection, we may now consider the full evolution of a PBH mass and spin taking into account the effects of both superradiance and Hawking evaporation, given by:
\begin{equation}
    \frac{dM}{dt} = -\mathcal{F}(\tilde{a}){M_P^4\over M^2} -\mu\Gamma_s N ~,
    \label{dMdt}
\end{equation}
\begin{equation}
    \frac{d\tilde{a}}{dt}= \tilde{a}{M_P^4\over M^3}(-\mathcal{G}(\tilde{a}) + 2 \mathcal{F}(\tilde{a}))-{M_P^2\over M^2}\Gamma_s N~.
    \label{dadt}
\end{equation}

\begin{figure}[htbp]
    \centering
    \includegraphics[scale=0.8]{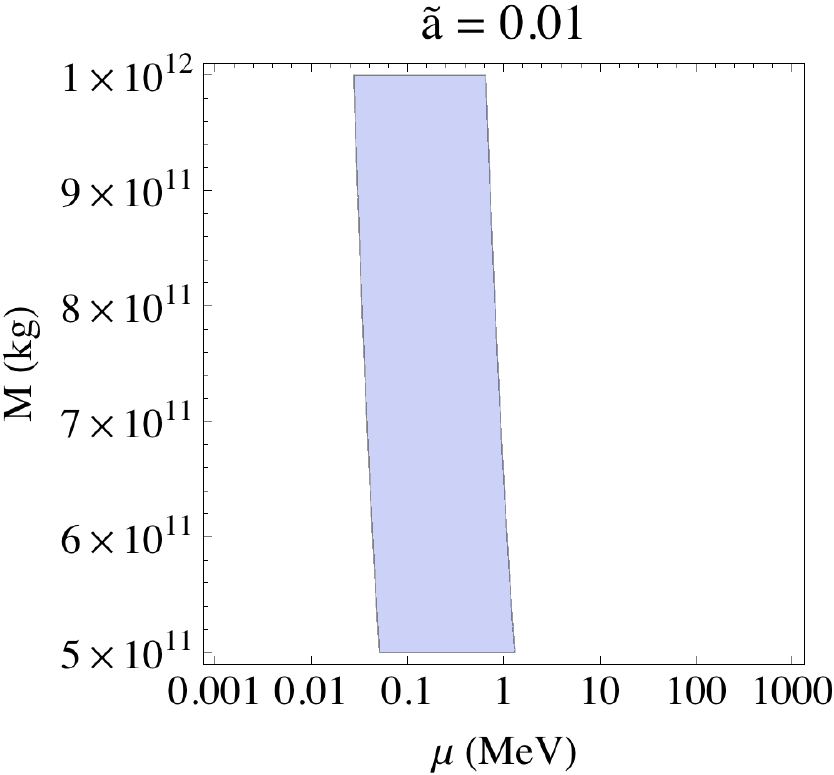}
    \caption{Value of the axion mass for which superradiant instabilities are triggered at PBH formation, for PBHs with a lifetime comparable to the age of the Universe born with $\tilde{a}_i=0.01$.}
    \label{sup_initial}
\end{figure}

As previously discussed, we are interested in PBHs with a lifetime close to the age of the Universe, i.e. with an initial mass in the range $5\times10^{11}-10^{12}$ kg, and particularly those born in the radiation era, with initial spins at or below the percent level. Despite their low spins, such PBHs may be superradiantly unstable already at formation, provided there are axions within the string axiverse in the right mass range. In particular, for PBHs with such mass and spin, superradiant instabilities may be triggered for axions with mass $\mu \lesssim 1$ MeV, but the axion mass cannot be too low, since the instability growth rate is proportional to $\mu^9$ as given approximately in Eq.~(\ref{Gamma_s}). Note, furthermore, that a significant amount of spin is only extracted from the PBH once the number of particles within the superradiant cloud $N\sim \tilde{a}M^2/M_P^2\sim 10^{37}(\tilde{a}/0.01)(M/10^{12}\ \mathrm{kg})^2$, requiring $\mathcal{O}(100)$ e-folds of superradiant amplification. This means that superradiance is only efficient for axions roughly in the 0.1-1 MeV mass range for PBHs born in the radiation-era, as illustrated in Fig.~\ref{sup_initial}.

Although there may be string theory compactifications including one or possibly more axions in this mass range, this is certainly not a generic expectation, since axion masses are exponentially sensitive to the magnitude of the non-perturbative effects that generate them. The hundreds or even thousands of light axions expected in realistic string compactifications should have masses distributed throughout a wide range of mass scales. Hence, scenarios with an axion in the mass range shown in Fig.~\ref{sup_initial} are certainly possibly but not necessarily the most likely, so we will focus our discussion henceforth in scenarios where nearly all axions have masses well below the MeV scale (contributing to the Hawking emission spectrum already at PBH formation), with possibly one extra axion above the MeV scale. The latter will not contribute to the initial Hawking spectrum (although it will once the PBH becomes hot enough), nor will it be produced via the superradiant instability until the PBH spin increases sufficiently as a result of evaporation.

To better understand the dynamical interplay between evaporation and superradiance, we start by considering a toy model where a PBH evaporates through the emission of a single light axion (well below the MeV mass scale), while superradiant instabilities may be triggered for a heavy axion of mass $\mu\gg$ 1 MeV. Although unrealistic, this will help us identifying the main qualitative features of the problem without the intricacies of adding the Standard Model particles across different mass thresholds. 

\begin{figure}[htbp]
    \centering
    \includegraphics[width=0.50\textwidth]{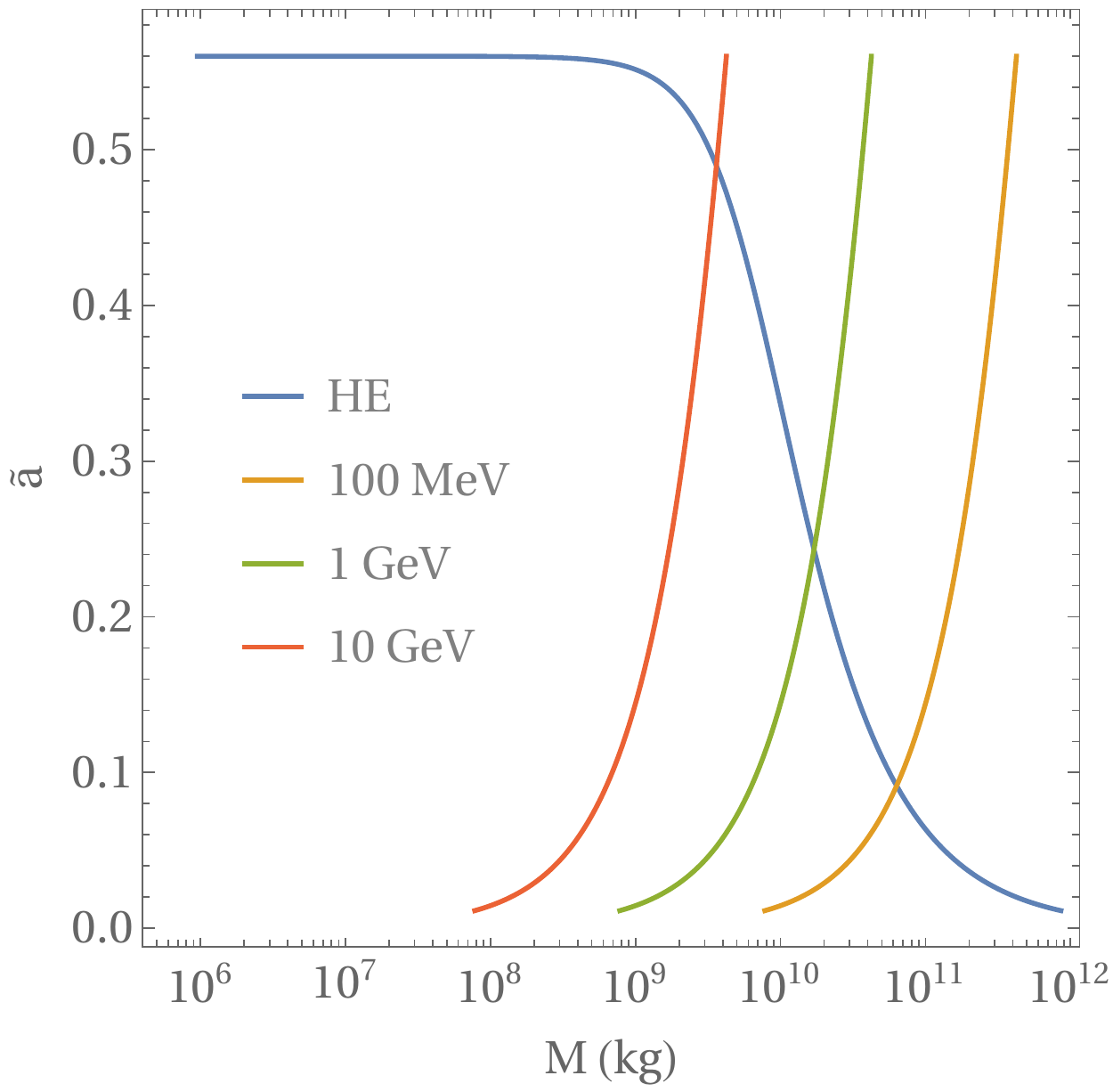}
    \caption{PBH Regge trajectory for single-scalar Hawking emission (HE) for $M_i=10^{12}$ kg and $\tilde{a}_i=0.01$, and superradiance threshold curves for different heavy axion masses, as labelled.} 
    \label{contour}
\end{figure}

In Fig.~\ref{contour} we show the PBH spin as a function of its mass considering only the effects of single scalar Hawking emission, obtained by solving numerically Eqs.~(\ref{dMdt}) and (\ref{dadt}) for an initial PBH mass $M_i=10^{12}$ kg and spin $\tilde{a}_i=0.01$. In this figure we also give the curves in the PBH mass-spin plane corresponding to the superradiance threshold $\omega=\Omega_H$ for different heavy axion masses.

The dynamics is thus expected to develop as follows. Initially, while the PBH spin is below the superradiance threshold for a given heavy axion mass, the latter is non-superradiant and any quantum fluctuations in the corresponding field are damped by the PBH. However, light scalar emission through the Hawking effect increases the PBH spin until at some point it crosses the threshold for superradiant heavy axion production. Any subsequent quantum fluctuation in the heavy axion field is then expected to be exponentially amplified via the superradiant instability, leading to the growth of a heavy axion cloud around the PBH.

\begin{figure}[htbp]
    \centering
    \includegraphics[width=0.50\textwidth]{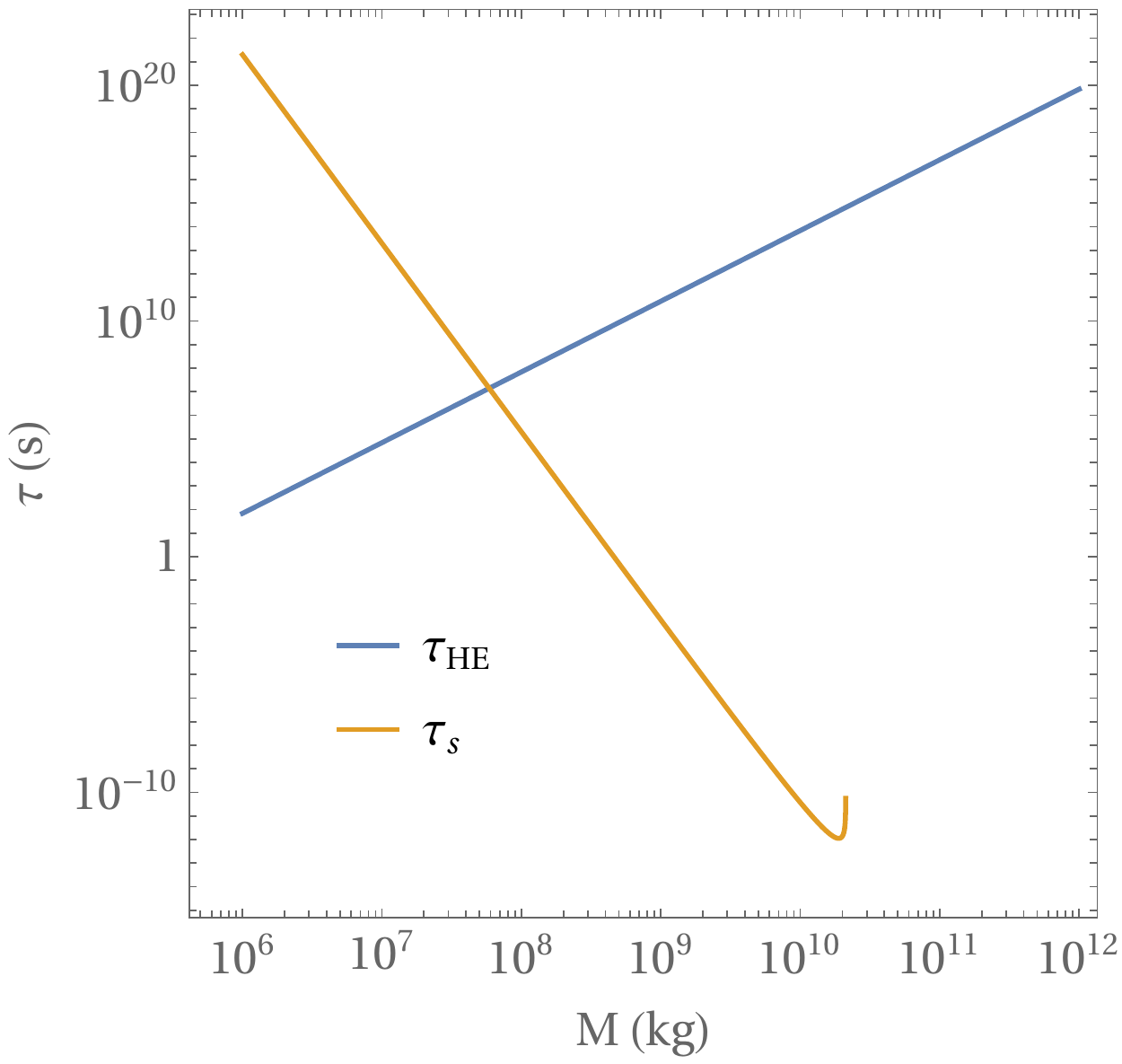}
    \caption{Characteristic timescales for Hawking emission (HE), $\tau_{HE}=\mathcal{F}^{-1}M^3/M_P^4$, and superradiance, $\tau_s=1/\Gamma_s$, for a PBH with $\tilde{a}=0.3$ and a heavy axion with mass $\mu=100$ MeV, as a function of the PBH mass.}
    \label{timescalesfig}
\end{figure}

We note that the timescales for superradiance and Hawking emission above the threshold differ by several orders of magnitude. For instance, as one can see in Fig.~\ref{contour}, for $\mu\sim 100$ MeV the superradiance threshold is attained when $\tilde{a}\sim 0.1$ and $M\sim 10^{11}$ kg. Such a PBH will evaporate in $\sim 10^8$ years ($\mathcal{F}(\tilde{a}=0.1)\sim 10^{-4}$), while the superradiance e-folding time when e.g.~the spin exceeds the critical value by 1$\%$ is $\sim 10^{-14}$ s. We illustrate this in Fig.~\ref{timescalesfig}, where we plot the Hawking evaporation and superradiance timescales for a heavy axion with $\mu=100$ MeV and a given PBH spin, as a function of the PBH mass.

As one can see in this figure, superradiance is a much faster process for the larger values of the PBH mass, implying that this will be the dominant process determining the PBH mass and spin after the critical spin value yielding $\omega<\Omega_H$ is reached. 
This difference in the timescales of the two processes may pose a numerical challenge for solving Eqs.~(\ref{dMdt}) and (\ref{dadt}) alongside $dN/dt=\Gamma_sN$ for the number of particles within the superradiant cloud. Nevertheless, we have found that the numerical tools available in e.g.~Mathematica are sufficiently accurate for this purpose. An alternative possibility is to artificially reduce the superradiant growth rate via a tunable multiplicative factor and then extrapolate the obtained results to the realistic case. We find that these two methodologies yield results consistent with each other.

A further numerical difficulty is crossing the superradiance threshold, since $N$ decreases exponentially fast in the non-superradiant regime, thus quickly reaching values below numerical precision before the PBH attains the critical spin value through light scalar Hawking emission. This, however, does not correspond to a realistic approach, since it discards the quantum nature of the heavy axion field. Although the development of superradiant instabilities from quantum field fluctuations has not, to our knowledge, been studied in detail so far, it is widely believed that superradiance will amplify any quantum field fluctuations, quickly increasing the corresponding occupation number in the quasi-bound state, so that a classical description is then sufficient to describe the dynamics. 

In fact, Kofman showed \cite{Kofman:1982gu} that Hawking emission populates not only free states, with $\omega>\mu$, but also quasi-bound states $\omega<\mu$, in a semi-classical calculation similar to the original computation by Hawking. While Kofman's analysis considered only a static BH, so that bound particles produced by Hawking emission are quickly reabsorbed by the BH, in principle it should extend also to the rotating case. The difference for a Kerr BH should reside in the exponential amplification of the bound state occupation number for spin parameters above the superradiance threshold.

\begin{figure}[htbp]
	\centering
	\includegraphics[width=0.50\textwidth]{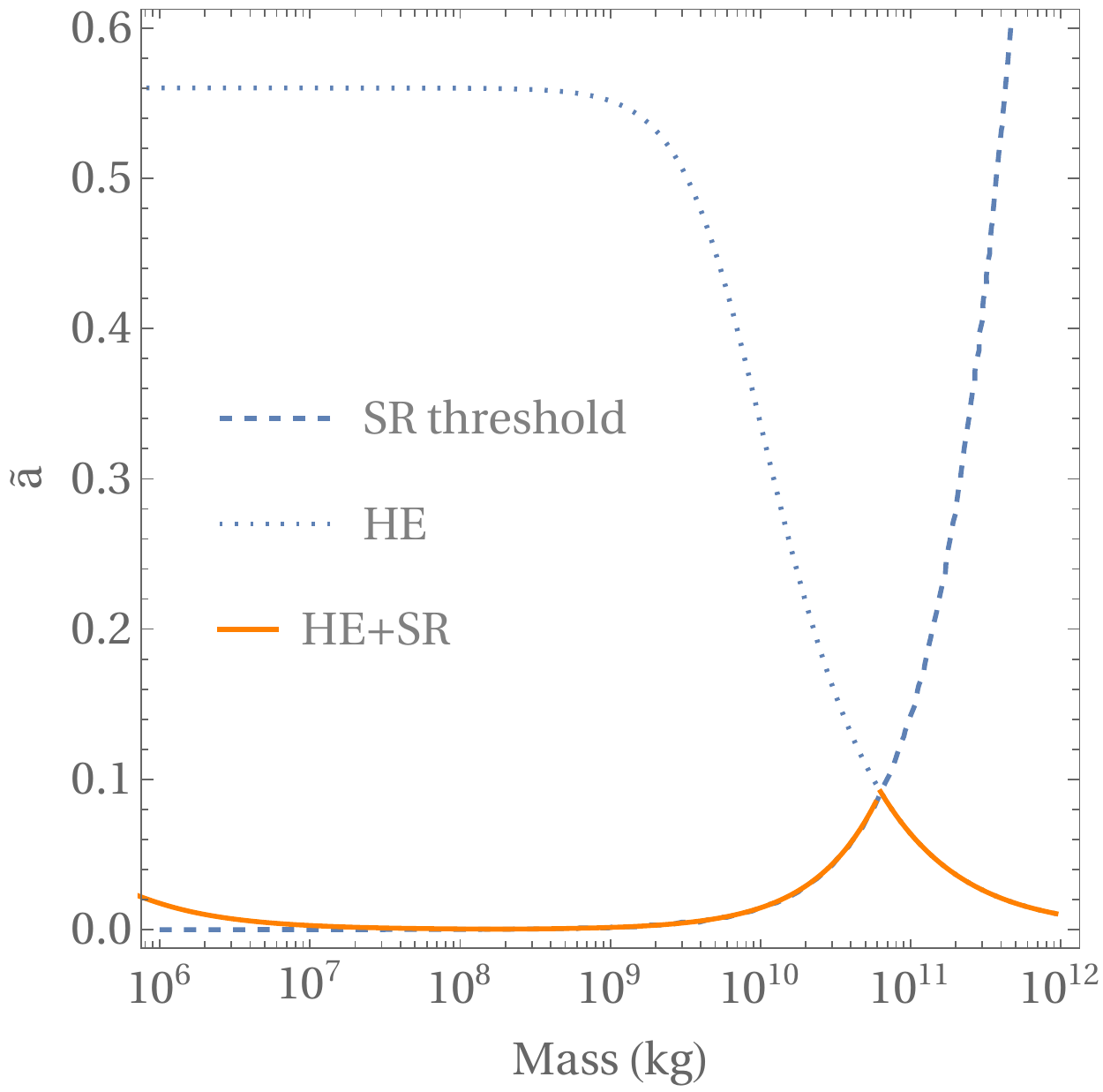}
	\caption{Black hole trajectory in the mass-spin ``Regge'' plane through Hawking emission and superradiance (solid orange curve) for $M_i=10^{12}$ kg, $\tilde{a}_i=0.01$ and a heavy axion with mass $\mu=100$ MeV. Also shown are the trajectory in the absence of superradiance (dotted blue curve) and the superradiance threshold (dashed blue curve).}
	\label{scalarmoney}
\end{figure}

In our numerical analysis, we assume this to be the case, and we simulate the effect of bound state quantum emission by first setting $N=0$ in the differential equations for the PBH mass and spin evolution, Eqs.~(\ref{dMdt}) and (\ref{dadt}), until just after the superradiance threshold is crossed within our numerical precision. We then take the obtained mass and spin values as initial conditions for the subsequent evolution, where we include the heavy axion cloud starting with $N=1$ (changing this initial value somewhat does not significantly affect our results). In the example shown in Fig.~\ref{scalarmoney}, we begin with $M_i=10^{12}$ kg and $\tilde{a}=0.01$, while the second part of the simulation including a heavy axion with $\mu=100$ MeV starts with $M = 6.30\times10^{10}$ kg and $\tilde{a} = 0.097$.

\begin{figure}[htbp]
	\includegraphics[width=0.45\textwidth]{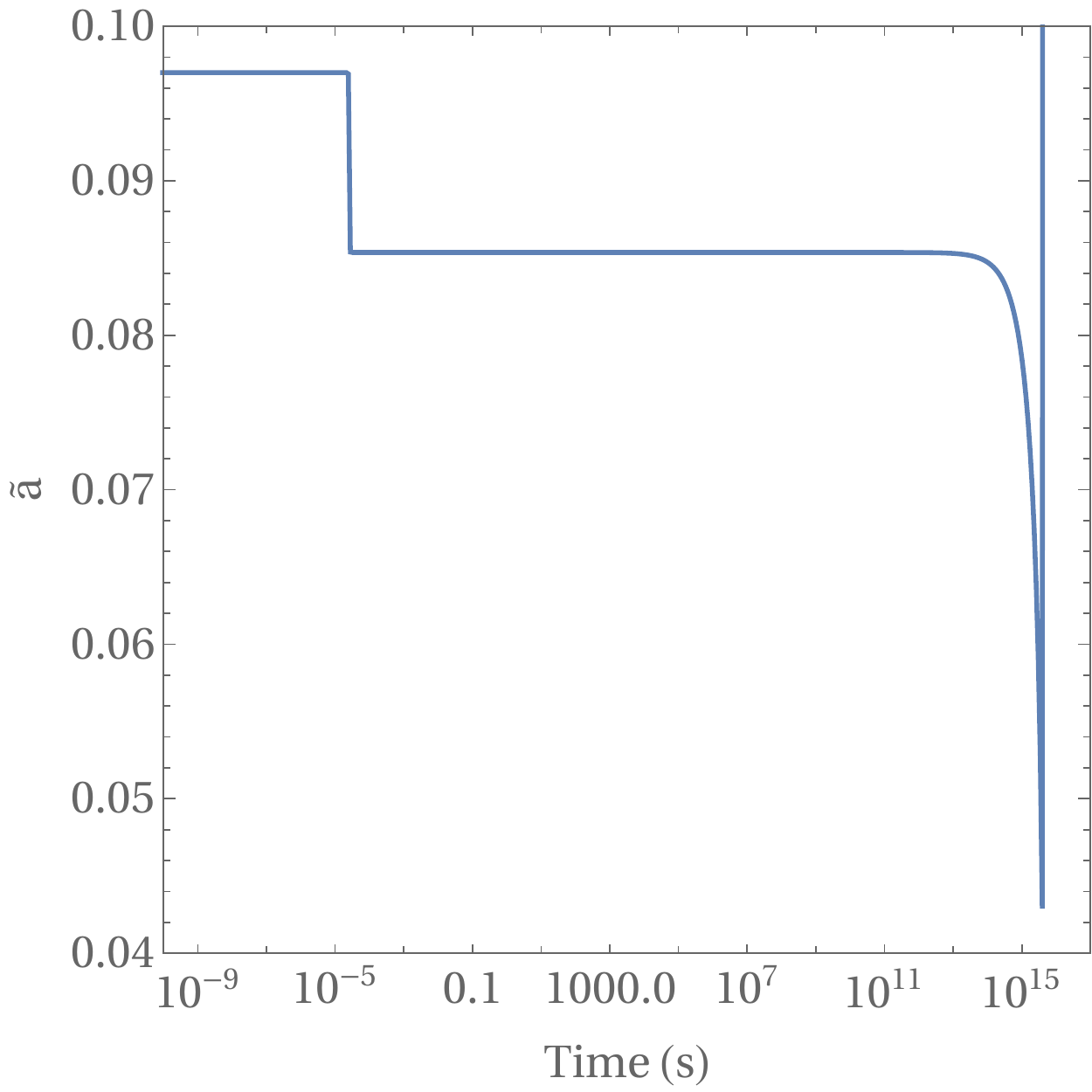}\hspace{0.7cm}
	\includegraphics[width=0.465\textwidth]{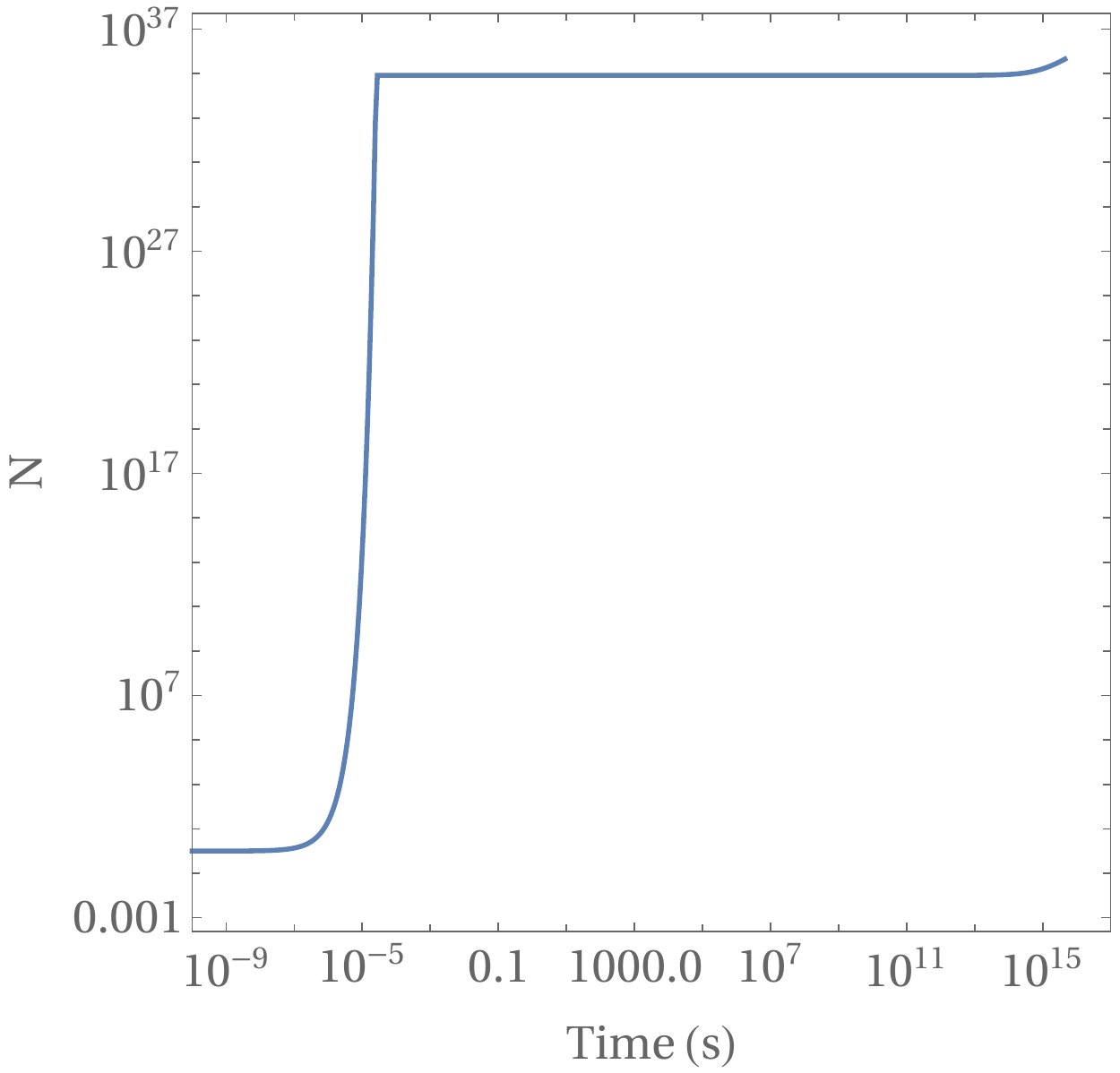}
	\caption{Numerical evolution of the PBH spin (left) and number of heavy axions in the superradiant cloud (right) for the same parameters of the example shown in Fig.~\ref{scalarmoney}. In these plots time is measured from the onset of the superradiant instability. }
	\label{scalar_N_a}
\end{figure}

As one can see in this figure, once the superradiant instability is triggered after the critical spin value is attained, the PBH follows closely the superradiance threshold. This is simply due to the fact that the latter occurs on much shorter timescales, quickly depleting the PBH spin until $\omega=\Omega_H$ and superradiant heavy axion production is halted. However, this condition is never fully attained since Hawking emission continuously spins up the PBH due to light axion emission. To better illustrate this, we show in Fig.~\ref{scalar_N_a} the time evolution of the PBH spin parameter and of the number of heavy axions in the superradiant cloud for the same example.

This shows that the number of heavy axions produced by superradiance grows exponentially fast after the instability is triggered, quickly decreasing the PBH spin back to close to the critical value. As one can observe in Fig.~\ref{scalar_N_a}, this does not constitute a very significant decrease in the PBH spin, since the superradiant instability is triggered just above the critical value at $\tilde{a}\simeq 4\alpha$. In fact, the number of heavy axions increases only until the superradiant term in Eq.~(\ref{dadt}) becomes comparable to the Hawking emission term. At this stage the system reaches a quasi-equilibrium, in which the spin-down effect of superradiance is nearly compensated by the spin-up due to Hawking evaporation. Setting $d\tilde{a}/dt\simeq 0$ and $\tilde{a}\simeq 4\alpha$ in Eq.~(\ref{dadt}) then yields the quasi-equilibrium condition:
\begin{equation}
    \centering
    \Gamma_s = \frac{4\mu(2 \mathcal{F}(\tilde{a})-\mathcal{G}(\tilde{a}))}{N}
    \label{eqcondition}
\end{equation}
This is analogous to the condition found in \cite{March-Russell:2022zll}, although in the latter case the opposite effect was observed since, in the absence of scalar emission, Hawking evaporation tends to spin down the PBH, leading to a reabsorption of the (initially superradiant) cloud in the $\Gamma_s<0$ regime. In the present case the cloud remains in the superradiant regime, i.e.~with $\Gamma_s>0$, so that superradiance produces more and more heavy axions within the cloud as evaporation continues to spin up the PBH. Since the product $\Gamma_s N$ is approximately constant, the number of particles grows linearly in this phase at a rate $4\mu(2\mathcal{F}-\mathcal{G}))\simeq 4\times 10^{19}$ axions per second in this example.

This quasi-equilibrium configuration is maintained only while the number of heavy axions within the superradiant cloud does change significantly, in this example up until $\sim 10^{13}$ s. After this, superradiance efficiently spins down the PBH, keeping the spin parameter very close to the critical value.

Despite the large number of heavy axions produced until this stage, superradiance has little effect on the PBH mass, which only begins to decrease after $\sim 10^{15}$ s ($\sim 30$ Myrs), corresponding to the remaining lifetime of the PBH when the superradiant cloud forms. 

The subsequent decrease in the PBH mass has two important effects, since it decreases the dimensionless mass coupling $\alpha=\mu M/M_P^2$. First, it lowers the critical spin value for which $\omega=\Omega_H$; second, it damps the superradiance growth rate $\Gamma_s\propto \alpha^8$. The first effect makes the PBH follow a trajectory in the Regge plane corresponding to the superradiance threshold, as observed in Fig.~\ref{scalarmoney}. This holds while superradiance remains faster than evaporation despite the decreasing PBH mass, i.e.~down to masses $\sim 10^7$ kg. This means that in its final hour (literally in this example) the PBH spins up once more as light scalar emission takes over in the last stages of evaporation. Asymptotically the PBH reaches the stable value $\tilde{a}_*=0.555$ yielding $\mathcal{H}(\tilde{a}_*)=0$ for pure scalar Hawking emission, as discussed in Section 2. This is not visible in Fig.~\ref{scalarmoney}, since it is only attained in the very last stages of the PBH evaporation, beyond the reach of the numerical precision of our simulation.

To summarize our findings in this toy model, a PBH formed with a mass $\sim 10^{12}$ kg evaporates through light axion emission and consequently spins up. After nearly $\sim 14$ billion years, its spin surpasses the critical value for triggering a superradiant instability, producing a cloud of heavy axions around it. For most of its remaining lifetime, the PBH is in a quasi-equilibrium configuration with the heavy axion cloud, with evaporation spinning up the PBH nearly at the same rate superradiance spins it down. In our working example the PBH remains in this stage for about 30 million years. At the end of its life, its mass starts decreasing and the PBH follows a Regge trajectory along the superradiance threshold up until its very last stages where evaporation once more increases its spin.

We note that once superradiance becomes inefficient the number of heavy axions within the superradiant cloud stabilizes near the maximum possible value:
\begin{equation}
N_\mathrm{max}\simeq \tilde{a}_c\left({M_c\over M_P}\right)^2~,    
\end{equation}
where the subscript `c' indicates the PBH mass and spin parameter when superradiance is triggered. This corresponds to converting most of the PBH's angular momentum into heavy axions via the superradiant instability (but fueled by the spin up produced by light scalar Hawking emission). In our example this yields nearly $10^{36}$ axions. 

Although the critical PBH-mass spin values for superradiance change for different values of the heavy axion mass, we observe the same qualitative behaviour for all $\mu>$ few MeV (recalling that in the $0.1-1$ MeV mass range superradiance is triggered at PBH formation for $\tilde{a}=0.01$ as discussed earlier).

Our toy model should be an accurate description when the PBH can emit $N_a\gg 1$ light axions, up to an overall rescaling of the PBH lifetime by a factor $\sim N_a^{1/3}$.

\subsection{Realistic string axiverse scenarios}

With the basic understanding of the main dynamical features of the interplay between superradiance and evaporation in the simplified toy model, we now perform more realistic simulations, with a finite number $N_a$ of light axions in the Hawking emission spectrum alongside all the Standard Model degrees of freedom. As described in Section 2, each particle species is included in the emission spectrum once the Hawking temperature exceeds its mass (or effective mass as in the case of quarks and gluons above the QCD scale).

\begin{figure}[htbp]
	\centering
	\includegraphics[width=0.5\textwidth]{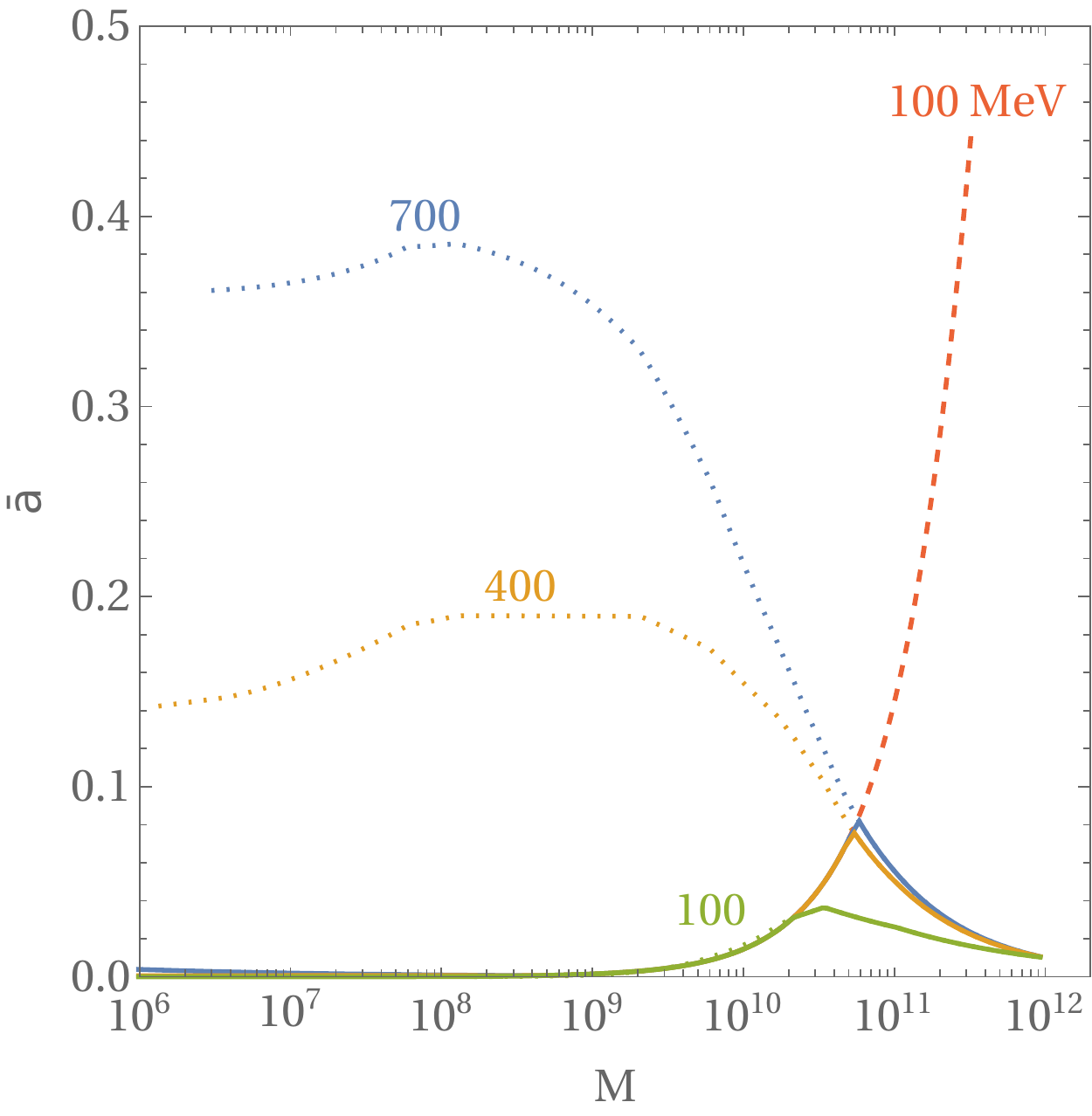}
	\caption{Black hole trajectory in the mass-spin ``Regge'' plane through Hawking emission and superradiance (solid curves) for $M_i=10^{12}$ kg, $\tilde{a}_i=0.01$ and a heavy axion with mass $\mu=100$ MeV, for different values of the number of light axions $N_a$ in the Hawking emission (HE) spectrum, as labelled. Also shown are the trajectory in the absence of superradiance (dotted curves) and the superradiance threshold (dashed curve).}
	\label{1scalarmoney}
\end{figure}

Given our understanding of the evaporation dynamics in the absence of superradiance, the main difference expected between the toy model and more realistic scenarios is the fact that most Standard Model particles have a non-zero spin, therefore carrying away part of the angular momentum of the BH. This means that Hawking emission is overall less efficient in spinning up the BH, and unless the number of light axions is sufficiently large the BH may actually spin down, as discussed in Section 2. The expectation is therefore that superradiant instabilities can only be triggered above a minimum number of light axion species $N_a$. This is illustrated in Fig.~\ref{1scalarmoney}, where we show the results of our numerical simulations for different numbers of light axions and a heavy axion with $\mu=$ 100 MeV.

As one can see in this figure, for different numbers of light axions the superradiant instability is triggered for different values of the PBH mass and spin, although converging to those found in the toy model in the limit $N_a\rightarrow \infty$. For $N_a\lesssim 100$ the superradiant threshold is not crossed for heavy axions with $\mu\gtrsim 10$ MeV, but since the string axiverse generically predicts hundreds or even thousands of light axions we typically expect instabilities to occur during the PBH evolution if axions in this mass range exist.

\begin{figure}[htbp]
	\includegraphics[width=0.45\textwidth]{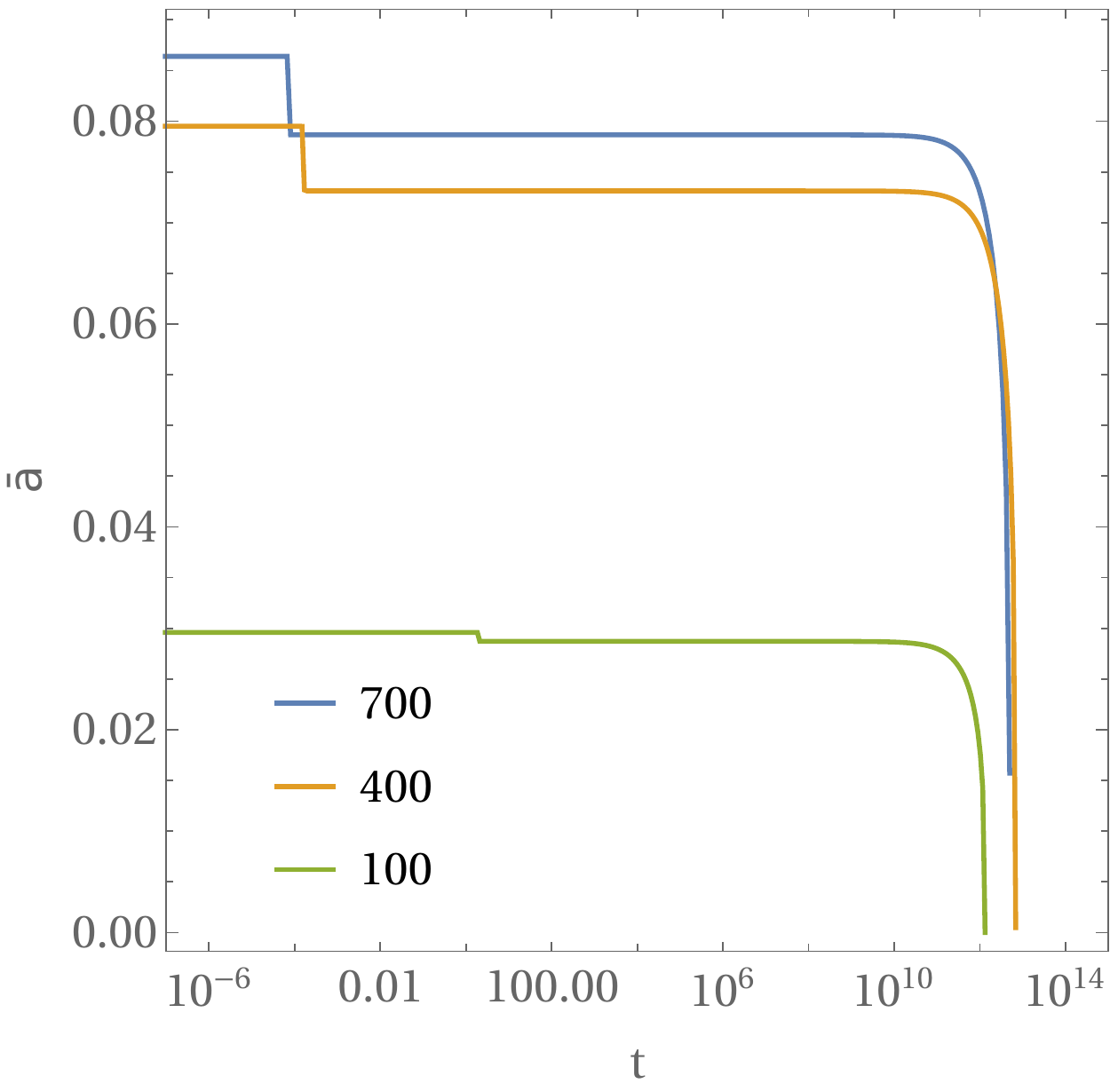}\hspace{0.7cm}
 \includegraphics[width=0.46\textwidth]{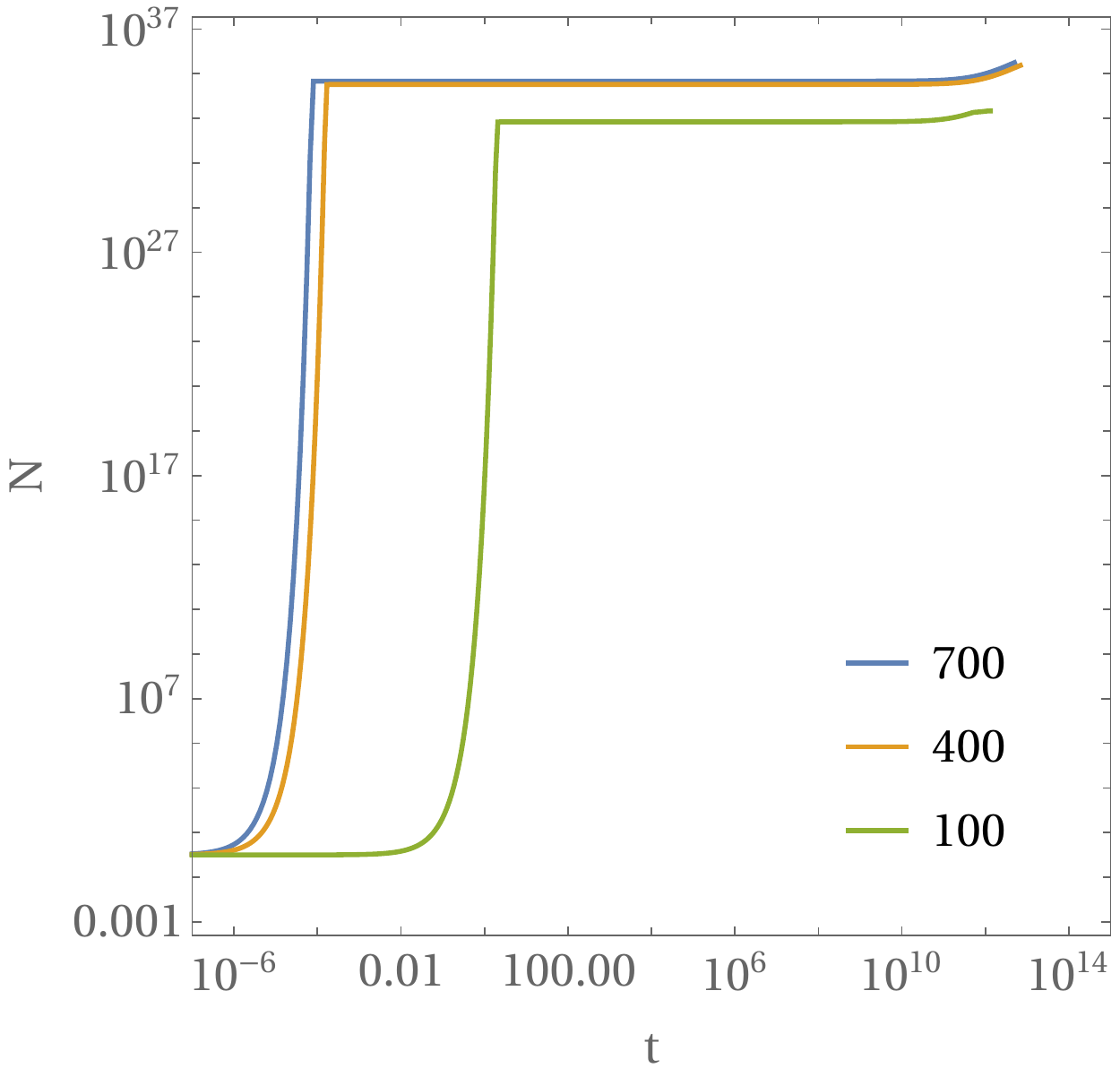}
	\caption{Numerical evolution of the PBH spin (left) and number of heavy axions in the superradiant cloud (right) for the same parameters of the example shown in Fig.~\ref{1scalarmoney}. In these plots time is measured from the onset of the superradiant instability.}
	\label{1at}
\end{figure}

As for the toy model, superradiance is initially much faster than Hawking emission in changing the PBH spin, so that after the instability is triggered the PBH follows a trajectory in the Regge plane corresponding to the superradiance threshold $\omega=\Omega_H$ ($\tilde{a}\simeq 4\alpha = 4 \mu M/M_P^2$ for slowly rotating PBHs). The main difference in realistic scenarios is the fact that we do not observe a spin up of the PBH for low masses, i.e.~at the end of its lifetime, as also clear in the time evolution plots shown in Fig.~\ref{1at},  given that Hawking emission is in this case much less efficient in increasing $\tilde{a}$ than for single scalar emission. 

\begin{figure}[htbp]
    \centering
    \includegraphics[scale=0.7]{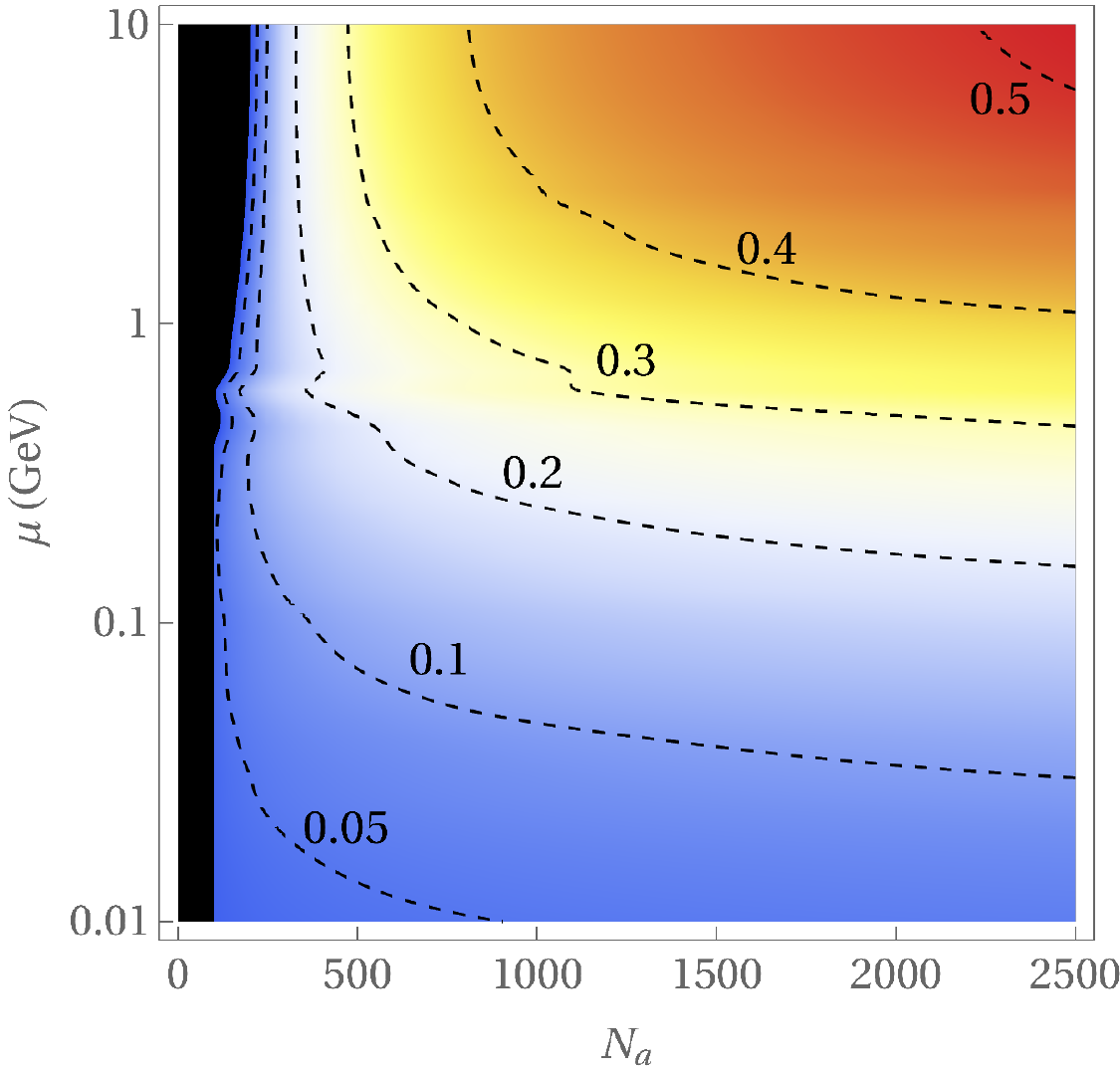}
    \caption{Contours of the critical spin parameter $\tilde{a}_c$ at which the superradiant instability is triggered as a function of the number of light axions $N_a$ and the mass of the heavy axion $\mu$. This corresponds to the present maximum spin of PBHs in the range $10^6-10^{12}$ kg (slowly rotating at birth). In the black region no superradiant instabilities are triggered.}
    \label{critical_spin}
\end{figure}

Although this may occur when the PBH reaches masses below those that our numerical precision can probe, we may safely conclude that for present PBH masses $\gtrsim 10^6$ kg (lifetime exceeding $\sim$1 s), the PBH distribution in the mass-spin Regge plane should exhibit a single peak at the values $(M_c,\tilde{a}_c)$ at which the instability is triggered and which depend on the string axiverse parameters $\mu$ and $N_a$. In particular, the mass of the heavy axion can be inferred from the superradiance threshold condition:
\begin{equation}\label{mu_relation}
\mu \simeq {M_P^2\over M_c}{\tilde{a}_c\over 2(1+\sqrt{1-\tilde{a}_c^2})}\simeq     {M_P^2\over 4 M_c}\tilde{a}_c~.
\end{equation}
The dependence on the number of light axions, $N_a$, emitted through the Hawking process is less trivial since it depends on the PBH evaporation dynamics, which has to be computed numerically. In Fig.~\ref{critical_spin} we show the critical spin contours in the $(\mu, N_a)$ plane, from which one can determine $N_a$ upon computing $\mu$ from Eq.~(\ref{mu_relation}).

We thus find a very unique signature of the string axiverse with hundreds of light ($\lesssim $0.1 MeV) axions and a single heavy axion ($\gtrsim$ few MeV), corresponding to a sharply peaked spin distribution as a function of mass, with a nearly linear relation between PBH mass and spin for masses below the peak. Moreover, as shown above, the number of light axions and the mass of the heavy axion can be determined from the position of this peak in the Regge plane, so that the full PBH distribution need not be probed across many orders of magnitude in mass.

This shows that measuring the present mass-spin distribution of PBHs below $10^{12}$ kg may have a very significant impact on finding (or excluding) new physics. Methodologies to determine both the mass and spin of a PBH from its photon Hawking emission spectrum have been developed by two of us in \cite{Calza:2022ljw}. Although these may be challenging from the experimental perspective, since they require measuring the PBH photon spectrum close to the primary emission peak energy (where the photon flux is lower than for the secondary component at lower energies), they may be within the reach of future gamma-ray telescopes, as we discuss in section 5.

We note that, in the presence of multiple heavy axions ($>$ few MeV), the first instability to be triggered during the evolution of a PBH corresponds to the lightest of these. The growth of this first heavy axion superradiant cloud will quickly spin down the black hole close to the corresponding superradiant threshold, as we have observed. This will therefore inhibit superradiant instabilities for heavier axions (except in the last fractions of a second of a PBH's lifetime where evaporation may still spin up the PBH). Hence, the shape of the present PBH mass-spin distribution is determined only by the lightest of the heavy axions, being largely insensitive to the existence of other axions.

We also note that our distinction between light axions and the heavy axion refers to the Hawking temperature of $\sim 10^{12}$ kg PBHs at formation. As they evaporate towards their present day mass, the Hawking temperature of these PBHs increases, such that at some stage the heavy axion can also be efficiently emitted. Since we are considering scenarios with $N_a\gg 1$, the inclusion of one (or even a few) more axion(s) does not significantly change the dynamics, and for simplicity we have kept $N_a$ fixed throughout the numerical evolution of the PBH mass and spin.

In our numerical simulations we have considered only free axions, i.e.~we have neglected the effects of axion self-interactions, which have been analyzed in detail in \cite{Gruzinov:2016hcq, Baryakhtar:2020gao} and also \cite{Branco:2023frw} (see also \cite{Yoshino:2012kn, Yoshino:2015nsa, Omiya:2020vji,Ferraz:2020zgi, Omiya:2022gwu, Omiya:2022mwv}). The latter considered superradiant axion production around rotating PBHs, although heavier than the ones considered in the present work so that the effects of Hawking emission could be neglected. Axion self-interactions lead, in particular, to 2-2 scattering processes that populate other superradiant and non-superradiant levels in the ``gravitational atom'' corresponding to the spectrum of BH-axion quasi-bound states. Some axions are ``ionized'' in these processes, escaping the BH's gravitational potential, which slows down the growth of the dominant 2p-superradiant cloud and may, in fact, prevent its occupation number from growing beyond a maximum number. 

The results obtained in \cite{Baryakhtar:2020gao, Branco:2023frw} cannot be easily extrapolated to the case of PBHs with a lifetime comparable to the age of the Universe, given how significant a role we have found PBH evaporation to play in the development of superradiant clouds. We may, nevertheless, try to estimate the parametric regimes in which it is a good approximation to neglect the effects of axion self-interactions, based on the analyses of \cite{Baryakhtar:2020gao, Branco:2023frw}. Since we are mostly interested in the non-relativistic regime, we may consider the effects of the leading non-linear term in the axion potential in the resulting Schr\"odinger-like equation, which has the Gross-Pitaevskii form:
\begin{equation}
i{\partial\psi\over\partial t}=-{1\over 2\mu}\nabla^2\psi -{\alpha\over r}\psi -{\lambda\over 8\mu^2}|\psi|^2\psi~,
\end{equation}
where the axion field $\Phi=(\psi e^{-i\mu t}+\mathrm{c.c.})/\sqrt{2\mu}$ and $\lambda=\mu^2/f_a^2$, with $f_a$ denoting the axion decay constant. In the limit $\lambda\rightarrow 0$ this corresponds to a Schr\"odinger equation for a Coulomb-like potential, yielding a Hydrogen-like spectrum of (quasi-)bound states as previously discussed. Self-interactions may thus play an important role when the non-linear term becomes comparable to the energy eigenvalue of the linear Hamiltonian, i.e.~when $\lambda|\psi|^2\sim \alpha^2\mu^3$ for the 2p-state ($\Phi\sim \alpha f_a$). Since $|\psi|^2$ represents the axion number density, and the 2p-cloud has approximately a toroidal shape with volume $V_\mathrm{cloud}=50\pi^2/(\alpha\mu)^3$ (with $(\alpha\mu)^{-1}$ yielding the gravitational Bohr radius) \cite{Rosa:2017ury,  Branco:2023frw}, we conclude that self-interactions can be neglected for $N\lesssim 50\pi^2/(\alpha\lambda)$. We may then derive a lower bound on $f_a$ by taking the maximum number of heavy axions produced in the 2p-cloud when the superradiant instability is triggered by PBH evaporation, $N_\mathrm{max}\simeq \tilde{a}_c(M_c/M_P)^2$:
\begin{equation}
f_a \gtrsim {\alpha_c^{3/2}\tilde{a}_c^{1/2} \over \sqrt{50}\pi} M_P\sim 7\times 10^{14} \left({\tilde{a}_c\over 0.1}\right)^2\ \mathrm{GeV}   
\end{equation}
where we recall that the subscript `c' refers to the PBH parameters when the superradiant instability is triggered by evaporation, with $4\alpha_c\simeq \tilde{a}_c$ in the slowly rotating limit. Since $\tilde{a}_c\lesssim 0.5$, given that Hawking emission cannot spin up a PBH beyond this value (which is only achieved for pure light scalar emission), we conclude that for heavy axions with decay constants above the grand unification scale, $f_a\gtrsim 10^{16}$ GeV, we may safely neglect the effects of self-interactions in the development of superradiant instabilities. Such large decay constants are, in fact, generic for string axions (see e.g.~\cite{Arvanitaki:2009fg}), thus justifying the free-axion approximation in this context.

We note that our dynamical simulations are applicable to any heavy scalar field ($\mu\gtrsim 1$ MeV) and not only axion-like fields, but the above arguments show that only for very feeble self-interactions may the dynamical effects of the latter be neglected. For instance, neutral pions are similar to heavy axions but interact quite strongly, with $\lambda\simeq 1$, as already analyzed in detail in \cite{Ferraz:2020zgi}.

\section{Direct detection of superradiant axion clouds}

In the previous section we have shown that PBH evaporation in the string axiverse may trigger superradiant instabilities for heavy axions due to the emission of hundreds (or even thousands) of light scalar axions and the consequent spin up of (initially slowly-rotating) PBHs. In addition to the unique imprint this leaves on the present mass-spin distribution of PBHs with masses $\lesssim 10^{12}$ kg, the formation of superradiant clouds may leave a much more direct observational signature, since the produced axions decay into photon pairs. In particular, as we will now describe in detail, an evaporating PBH surrounded by a heavy axion cloud will emit photons as a result of both Hawking emission and heavy axion decay, yielding a unique spectrum.



Hawking emission leads to two types of photons in a PBH emission spectrum. Primary photons are directly emitted by the PBH with a nearly-thermal spectrum (up to the gray-body factors discussed in section 2) given by \cite{Page:1976df, Page:1976ki}:
\begin{equation}\label{prim}
{d^2N_{\gamma,P}\over dt dE_\gamma}={1\over 2\pi}\sum_{l,m}{\Gamma^1_{l,m}(\omega)\over e^{2\pi k/\kappa}\pm 1}~,
\end{equation}
where $\omega=E_\gamma$ is the mode frequency (see section 2). In addition, charged particles produced via the Hawking effect also emit photons as they travel away from the PBH, and additional photons also result from the decay of unstable particles like the neutral pion\footnote{The $N_a$ light axions emitted by the PBH also decay into photons, but their lifetime is so long that they typically decay far away from the PBH, so that we do not include their contribution to the Hawking photon emission spectrum.}. Such secondary photons are less energetic than their primary counterparts but may nevertheless dominate the emission spectrum at energies below the primary emission peak. 

Although the primary spectrum can be computed using semi-analytical methods (computing the gray-body factors numerically as described in Section 2), determining the secondary spectrum typically requires numerical methods of convoluting the primary emission rate for each particle species (analogous to Eq.~(\ref{prim})) with their corresponding photon emission rate. We have used the publicly available BlackHawk code \cite{Arbey:2019mbc,Arbey:2020yzj,Arbey:2021yke,Arbey:2021mbl} to compute both the primary and secondary emission spectra of PBHs with mass and spin satisfying the superradiance threshold condition $\omega=\Omega_H$, corresponding to the trajectory followed by a PBH after the formation of a heavy axion superradiant cloud of a given mass $\mu\gtrsim 1$ MeV. We have nevertheless checked that our semi-analytical calculation of the primary emission spectrum agrees with the results obtained using this code. 

The latest version of BlackHawk uses two well-known particle physics codes to compute the number of photons radiated by primary particles, namely Hazma \cite{Sjostrand:2007gs,Bierlich:2022pfr} for primary particle energies below a few GeV and PYTHIA \cite{Coogan:2019qpu} for energies $>5$ GeV. PYTHIA code may operate in an extended range via extrapolation tables, but as reported in \cite{Coogan:2020tuf} this may lead to unreliable spectra due to its failure in describing physical processes as the neutral pion decay, $\pi^0\rightarrow \gamma\gamma$ which should cause a symmetric emission peak centered at half of the pion's mass. We note that the primary emission peak corresponds to photon energies $\sim 5$ times the Hawking temperature. For this reason, and taking into account the limits of validity of Hazma and PYTHIA, we employ PYTHIA for PBH masses $M < 2.5\times10^{10}$ kg, while for  $M>2.5\times 10^{10}$ kg we use Hazma.

The heavy axions within the superradiant cloud decay into photon pairs with a rate (see e.g.~\cite{GrillidiCortona:2015jxo,Bauer:2017ris}):
\begin{equation}\label{Gamma}
   \Gamma_{a}=\frac{g^2_{a \gamma \gamma} \mu^3}{64 \pi} \;,\qquad 
   g_{a \gamma \gamma} =\frac{\alpha_{EM}}{2 \pi f_a} |\mathcal{C}_{a \gamma \gamma}| \;,
\end{equation}
where $\alpha_{EM}\simeq 1/137$ is the electromagnetic fine structure constant and  $\mathcal{C}_{a \gamma \gamma}=\mathcal{O}(1-10)$ is a model-dependent numerical factor (possibly reaching larger values in some axion models (see e.g.\cite{Agrawal:2018mkd}). We may write this as:
\begin{equation}\label{Gamma2}
   \Gamma_{a}\simeq 7\times 10^{-35} |\mathcal{C}_{a \gamma \gamma}|^2\left({\mu\over 100\ \mathrm{MeV}}\right)^3\left({10^{16}\ \mathrm{GeV}\over f_a}\right)^2\ \mathrm{eV}~.
\end{equation}
Note that for the heavier axions this may exceed the present Hubble rate $H_0\sim 10^{-33}$ eV, i.e.~yield axions with a lifetime shorter than the age of the Universe. However, the heavy axion cloud is only formed after $\simeq 14$ Gyrs, once evaporation spins up the PBH sufficiently to trigger the superradiant instability. It is easy to check that the axion decay rate is always smaller than the PBH evaporation rate when the cloud forms:
\begin{equation}\label{decay_evap}
 {\Gamma_a\over \Gamma_\mathrm{evap}}={\alpha_{EM}^2|\mathcal{C}_{a \gamma \gamma}|^2\over 256\pi^3 \mathcal{F}}\left({M_P\over f_a}\right)^2 \alpha_c^3 \simeq |\mathcal{C}_{a \gamma \gamma}|^2\left({10^{-2}\over \mathcal{F}}\right)\left({10^{16}\ \mathrm{GeV}\over f_a}\right)^2\alpha_c^3 \ll 1
\end{equation}
since the function characterizing the PBH mass loss rate (see Section 2) $\mathcal{F}\gtrsim 10^{-2}$ for $N_a\gtrsim 100$ and, as discussed in the previous section, $f_a\gtrsim 10^{16}$ GeV for string axions, taking also into account that superradiance is triggered for $\alpha_c<0.15$ in all axiverse scenarios. This means that axion decay does not play a significant role in the formation and evolution of the superradiant clouds. It may, however, yield an observable signal as we now show. The corresponding photon emission spectrum is given by:
\begin{equation}\label{dN/dt}
    \frac{d^2N_{\gamma,a}}{dtdE_\gamma} \simeq 2\Gamma_a N \delta\left(E_\gamma-{\mu\over2}\right) \simeq {2\Gamma_a N \over \sqrt{2\pi}\Delta E}e^{-{(E_\gamma-\mu/2)^2\over 2\Delta E^2}}
\end{equation}
since each of the two photons has approximately half of the axion rest energy (up to sub-leading gravitational binding energy corrections) and, in the last step, we have replace the monochromatic spectrum by a Gaussian function of width $\Delta E$ in order to take into account the effects of a detector's resolution. We then obtain for the maximum photon emission rate from the superradiant axion cloud (at $E_\gamma\simeq \mu/2$), considering the maximum number of axions produced in the evolution, as computed in the previous section:
\begin{equation} \label{axion_peak}
    \left.\frac{d^2N}{dt dE_\gamma}\right|_\mathrm{max} \simeq 1.5 \times 10^{18} \left|C_{a\gamma\gamma}\right|^2 \left({\tilde{a}_c\over 0.1}\right)^3\left({10^{16}\ \mathrm{GeV}\over f_a}\right)^2 \left({\Delta E\over E_\gamma}\right)^{-1}\ \mathrm{GeV^{-1}s^{-1}} 
\end{equation}

The energy of the axion line is always smaller than the peak of the primary photon emission spectrum of the PBH, since the latter occurs for $E_\gamma\simeq 5T_H \simeq 0.2\alpha^{-1}\mu\gtrsim \mu$. Hence, whether the axion line is detectable depends on the magnitude of the secondary photon emission spectrum from Hawking evaporation.

\begin{figure}[htbp]
\includegraphics[scale=0.625]{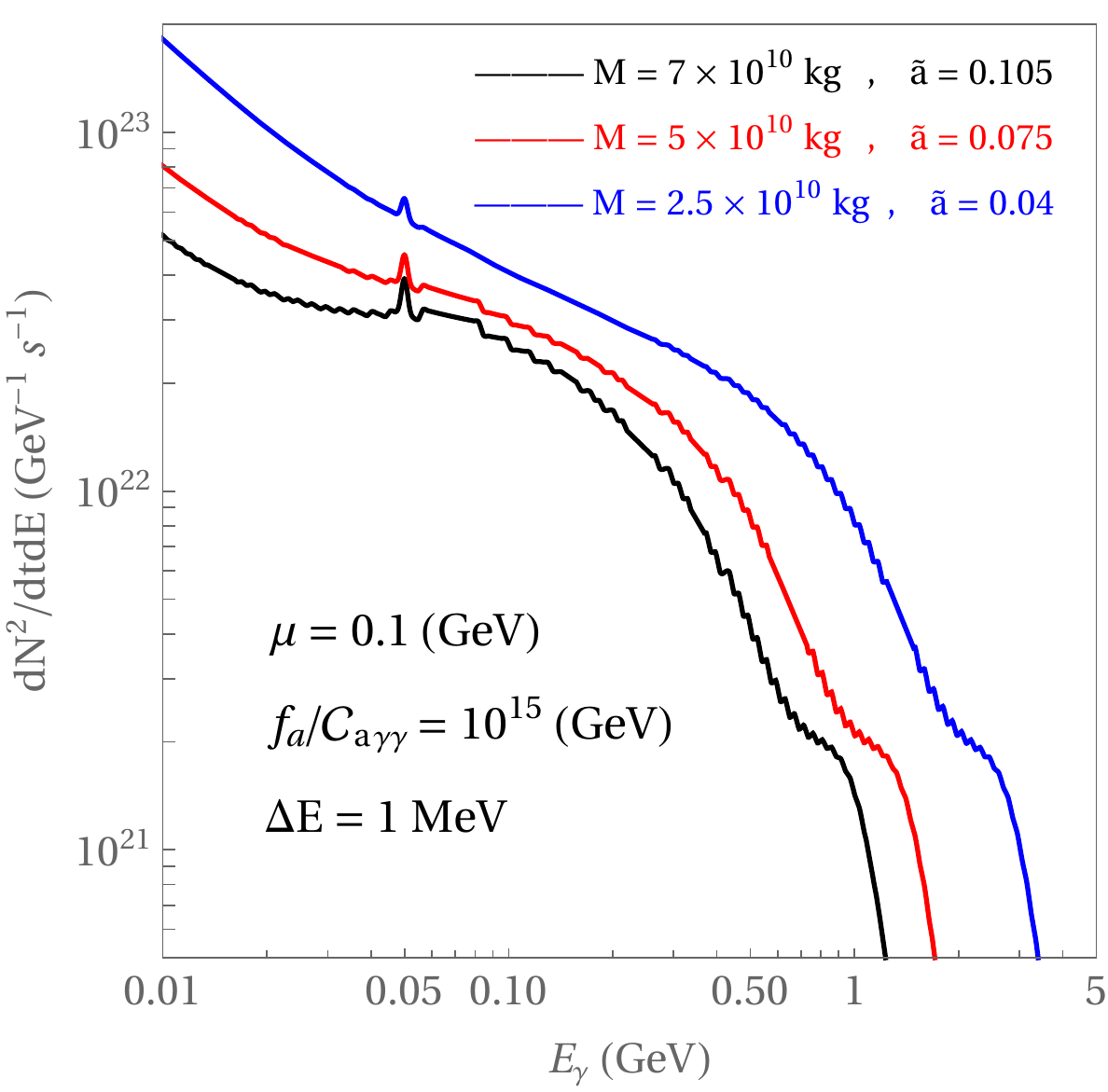}
\includegraphics[scale=0.62]{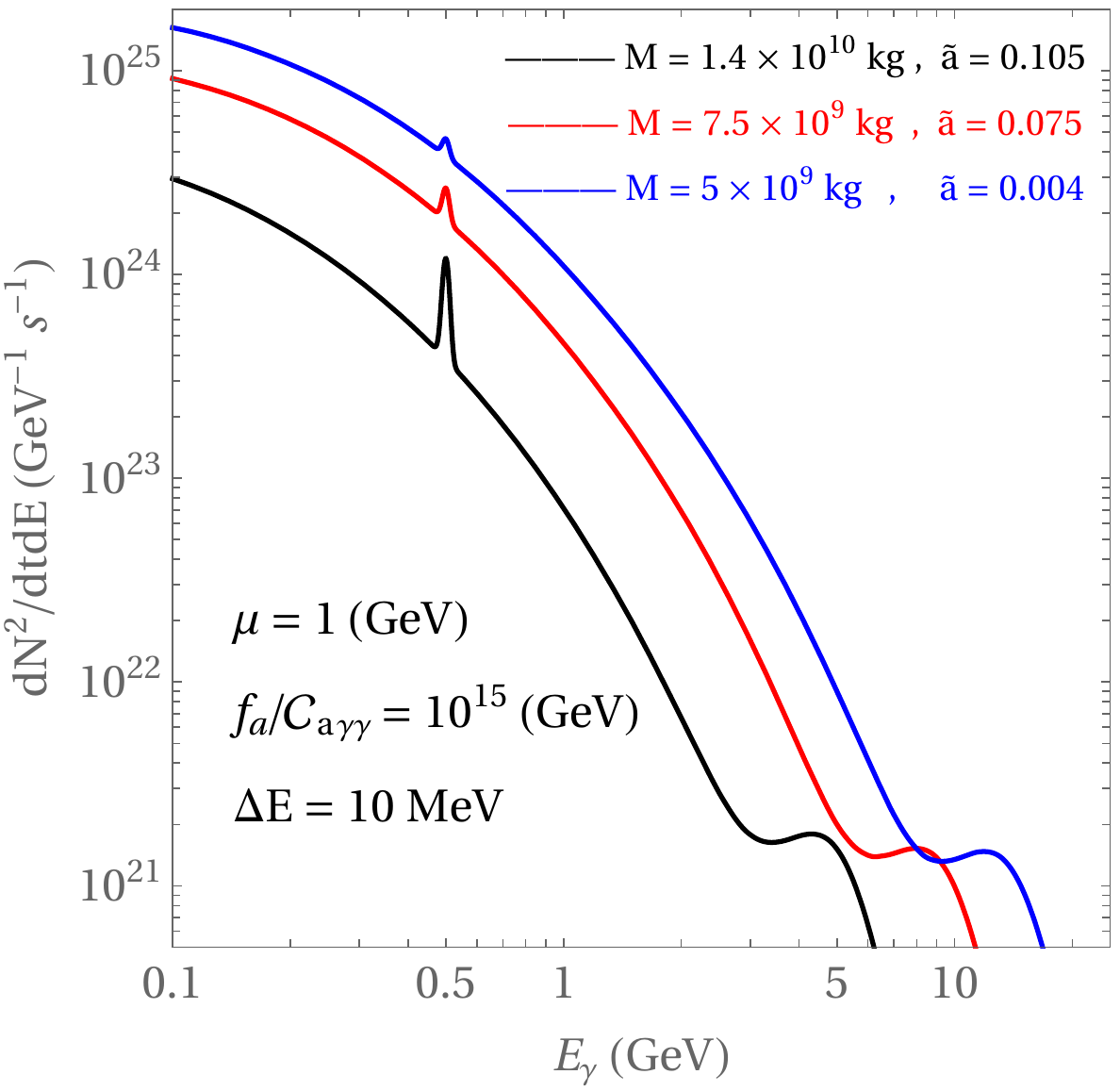}
\caption{Photon emission spectrum of PBHs in the presence of a heavy axion with 100 MeV (left) or 1 GeV (right) mass and  $f_a/\mathcal{C}_{a\gamma\gamma}=10^{15}$ GeV ($g_{a\gamma\gamma}\simeq 10^{-18}$ GeV). In each case the heaviest PBH (black curve) corresponds to the mass and spin values at which the superradiant cloud forms, while the red and blue curves correspond to subsequent stages in the latter's evolution (along the superradiance threshold curve). The energy resolution for the axion line is taken to be 1\% of the axion mass in each case.} 
\label{pbh_axion_spectrum}
\end{figure}

In Fig.~\ref{pbh_axion_spectrum} we give two examples illustrating the effect of a superradiant cloud with heavy axions with $\mu=100$ MeV and $1$ GeV on the emission spectrum of PBHs with three different masses and spin. The heaviest PBHs in each case correspond to a PBH where the heavy axion cloud has just formed (and reached its maximum mass), after evaporating with $N_a=400$ light axions for nearly the age of the Universe. The other two mass and spin values correspond to subsequent stages of the same PBH as it evaporates further and follows the Regge trajectory given by the superradiance threshold condition $\omega=\Omega_H$ ($\tilde{a}\simeq 4\alpha$) as discussed in the previous section. We note that in practice one would aim to observe three distinct PBHs presently at different stages of the evaporation process (already dressed with a heavy axion cloud), and not the same PBH at different times, since the evaporation timescale in this mass range is still very large ($\sim 1$ million years).

In these examples we have chosen $f_a=10^{16}$ GeV and $\left|C_{a\gamma\gamma}\right|=10$ (or equivalently any combination with $g_{a\gamma\gamma}\simeq 10^{-18}\ \mathrm{GeV}^{-1}$), which maximizes the intensity of the heavy axion line given the constraints obtained from neglecting axion self-interactions discussed above and the typical values of axion decay constants of string compactifications. We see that with a $\simeq 2$\% peak energy resolution the axion line is clearly visible above the secondary photon emission from the PBH evaporation for the PBH mass values considered. Although these examples may be somewhat optimistic, it is quite remarkable that such an axion line is observable for such low values of the axion-photon coupling. Note that in the case of heavier axions, for which the instability is triggered at higher spin values, the axion line is more pronounced as given by  Eq.~(\ref{axion_peak}).

It is worth remarking that detecting a slowly rotating black hole with a mass $\lesssim 10^{11}$ kg, which must in principle be a remnant of the evaporation of a heavier PBH\footnote{Although there could be more exotic scenarios where light black holes form much later (see e.g.\cite{Picker:2023ybp}).}, exhibiting a monochromatic line in its electromagnetic emission spectrum would be evidence for the existence not only of a heavy axion but also of hundreds of light axions, as otherwise it could not have developed a large enough spin to trigger the superradiant instability (recall that for $N_a=0$ any natal spin is quickly lost, as can be seen in Fig.~\ref{fig1}).

\section{Conclusions}

In this work we have considered the evaporation of PBHs in the context of the string axiverse, following on the seminal work in \cite{Calza:2021czr}. The generic prediction of hundreds or even thousands of light scalar axions in realistic string scenarios has a tremendous impact on the dynamics of small PBHs, since light scalar emission tends to spin up a BH, as opposed to the emission of particles with non-zero spin. This is due to spin zero particles being the only particles that can be emitted in the spherically symmetric $l=0$ mode, i.e.~without carrying away the BH's angular momentum. As shown in \cite{Calza:2021czr} and revised in detail in Section 2 of the present work, an initially slowly rotating PBH ($\tilde{a}\lesssim 0.01$) can spin up up to values $\tilde{a}\sim 0.1-0.5$ for $N_a\gtrsim 100$ light axions.

This increase in a PBH's angular velocity, which for PBHs born with $\sim10^{12}$ kg occurs on timescales comparable to the age of the Universe, has an important consequence that we have explored in detail in this work - it may trigger superradiant instabilities. The string axiverse typically includes axions with masses spread out over several orders of magnitude \cite{Arvanitaki:2009fg}, most of which are likely below the MeV scale and hence included in the PBH Hawking emission spectrum for the above-mentioned natal mass range. However, there may be one or more axions with a larger mass, and which can be produced via the superradiant instability once a PBH reaches a critical spin value as a result of evaporation.

The dynamical interplay between Hawking evaporation (with light sub-MeV axions) and the superradiant instability (producing heavy super-MeV axions in clouds gravitationally bound to the PBH) is quite interesting, given in particular the very different timescales of the two particle production processes. As we have shown in this work, once evaporation spins up a PBH above a certain critical spin, the superradiant instability quickly amplifies any quantum fluctuation in the heavy axion field, and the expense of reducing the PBH's spin back to the critical value. On a longer timescale, the PBH continues to spin up due to light axion emission, therefore feeding the superradiant instability and the heavy axion cloud. These two opposing effects keep the PBH-axion cloud system in a quasi-equilibrium state with nearly constant spin for a long time, and as the PBH mass decreases it follows a very simple Regge trajectory (mass-spin plane) corresponding to the superradiance threshold for the heavy axion $\tilde{a}\simeq4 \mu M/M_P^2$.

Towards the end of the PBH's lifetime superradiance becomes less and less efficient in extracting the PBH spin, as a consequence of the decreasing dimensionless mass coupling $\alpha=\mu M/M_P^2$. The number of heavy axions in the cloud stabilizes near the maximum value $N_\mathrm{max}\simeq \tilde{a}_c(M_c/M_P)^2$, where the subscript `c' denotes the PBH parameters when the superradiant instability is triggered, as supported by our numerical simulations. Evaporation then takes over as the main mechanism driving the PBH evolution and therefore increasing its spin for a sufficiently large number of light axions. Numerically, we can only observe this final spin up in the toy model with pure scalar Hawking emission, given that numerical precision limits the considered PBH mass range to $M_0\gtrsim 10^6$ kg\footnote{Such PBHs live less than a second, while our simulations span the age of the Universe, requiring a very large numerical precision.} in a toy model with pure scalar emission. This toy model mimics what happens in the limit $N_a\rightarrow \infty$.

This thus leads to a striking prediction for the present mass-spin distribution of PBHs in the range $10^{6}-10^{12}$ kg. On the one hand, for the heavier ones that are still spinning up due to light axion emission, the spin parameter should decrease with the mass (the exact function depending on the number of light axion species, $N_a$). On the other hand, for the lighter PBHs that have already formed a heavy axion cloud, the spin parameter should increase linearly with the PBH mass, along the Regge trajectory corresponding to the superradiance threshold $\tilde{a}\simeq4 \mu M /M_P^2$. This gives a peaked mass-spin distribution (see Fig.~\ref{1scalarmoney}), the mass and spin of the most rapidly rotating PBH depending on the number of light axions, $N_a$, and the mass of the heavy axion $\mu$ (see Fig.~\ref{critical_spin} and associated discussion).

In addition to this indirect signature of the string axiverse, the presence of a superradiant axion cloud can in principle be directly detected as a single emission line on top of the PBH's Hawking emission spectrum, located at approximately half of the heavy axion's mass (since axions decay into photon pairs). Although we have not performed a detailed analysis of the detectability of this axion line, we have shown that its intensity can be comparable to that of the PBH's (secondary) Hawking emission for axion-photon couplings as low as $g_{a\gamma\gamma}\sim 10^{-18}\ \mathrm{GeV}$, corresponding to axion decay constants of the order of the grand unification scale, $f_a\sim 10^{16}$ GeV, typical of string axions, up to an $\mathcal{O}(10)$ model-dependent coefficient $\mathcal{C}_{a\gamma\gamma}$. This feature is quite unique, since PBHs with different masses and spins, presently at distinct evaporation stages, should exhibit the same axion line despite their different Hawking emission spectra if they have grown a superradiant cloud around them.

Both indirect and indirect signatures of the string axiverse depend intrinsically on detecting and accurately measuring a PBH's photon emission spectrum, since in addition to the axion line this allows for a determination of both its mass and spin, following e.g.~the methodologies devised in \cite{Calza:2022ljw}. In particular, the latter require determining specific features in the spectrum close to the primary emission peak, where, as illustrated in Fig.~\ref{pbh_axion_spectrum} the emission rate is lower. As discussed in \cite{Calza:2022ljw}, the sensitivity of planned gamma-ray telescopes such as the All-Sky-ASTROGAM \cite{e-ASTROGAM:2016bph, Tatischeff:2019mun} or AMEGO \cite{AMEGO:2019gny, Fleischhack:2021mhc} missions may not be sufficient for these purposes unless we can find light PBHs ($\lesssim 10^{12}$ kg) at a distance below $100$ AU of the Earth, which although not impossible is unlikely given current bounds on their abundance \cite{Carr:2020gox}. However, the proposed MAST mission \cite{Dzhatdoev:2019kay}, with an unprecedentedly large detector area, may potentially reach enough sensitivity. 

An important question also comes out of our analysis in this work $-$ what happens to the heavy axion clouds once the PBHs evaporate away? The analysis of superradiant dark matter production by light PBHs ($<10^6$ kg) performed in \cite{March-Russell:2022zll} has suggested (although not rigorously proven), that superradiant clouds may survive black hole evaporation as self-gravitating, microscopic boson stars. The main idea is that, as a PBH evaporates, its gravitational potential (which bounds the scalar cloud) decreases in time, first adiabatically (compared to the timescale of the Hydrogen-like wave function), but speeding up towards the end of the PBH's lifetime so that the PBH suddenly vanishes - much like a quantum quench. Using the results obtained in \cite{March-Russell:2022zll}, we find that PBH evaporation should only become non-adiabatic when the PBH reaches a value:
\begin{equation}
M_*\simeq 7\times10^{-5}\left({0.1\over \tilde{a}_c}\right)^{13/5}\left({10^{11}\ \mathrm{kg}\over M_c}\right)^{2/5} M_\mathrm{cloud}~,
\end{equation}
where $M_\mathrm{cloud}=\mu N$ is the total mass of the axion cloud. This then suggests that the heavy axion field profile should slowly evolve from a superradiant cloud around a PBH to an essentially self-gravitating configuration well before the PBH fully evaporates away. PBH evaporation could thus leave behind microscopic axion stars! Note that the cloud expands from an initial size of a few times the gravitational Bohr radius $\sim M_P^2/M_c\mu^2$ to the much larger size of the self-gravitating configuration, $\sim$few$\times M_P^2/M_\mathrm{cloud}\mu^2$, given that the axion cloud only contains in general a small fraction of the PBH mass when it forms, i.e.~$M_\mathrm{cloud}\ll M_c$. Note also that this should result in a rotating boson star by angular momentum conservation, but that these configurations are unstable and end up decaying into non-rotating spherical stars \cite{Sanchis-Gual:2019ljs, DiGiovanni:2020ror, Dmitriev:2021utv}.

Showing that superradiant clouds may indeed become self-gravitating states requires dedicated numerical simulations, given the intrinsically non-linear nature of the problem, and which are beyond the scope of this work. Nevertheless, it is interesting to speculate about the possibility of directly observing such a transition, since after its final Hawking explosion, a PBH could leave behind a compact object (the ``axion star'')  with a monochromatic gamma-ray spectrum, as computed in the previous section.

Whether or not sufficiently sensitive telescopes will become available within the foreseeable future to detect all the effects proposed in this work, this demonstrates the enormous potential that evaporating PBHs can have as probes of beyond the Standard Model physics, in particular the string axiverse. We can only hope that the Universe has been kind enough to provide us with a sufficiently large number of these fascinating compact objects. 

\begin{acknowledgments}
M.C. is supported by the FCT doctoral grant SFRH/BD/146700/2019. This work was supported by national funds from FCT - Funda\c{c}\~ao para a Ci\^encia e a Tecnologia, I.P., within the project UID/04564/2020 and the grant No.~CERN/FIS-PAR/0027/2021.
\end{acknowledgments}


\end{document}